\documentclass[prd,showpacs,twocolumn,
amsmath,amssymb,superscriptaddress,floatfix,nofootinbib]{revtex4-1}
\usepackage[utf8]{inputenc}
\usepackage[sort&compress]{natbib}
\usepackage[normalem]{ulem}
\usepackage{lipsum} 
\usepackage{bm}
\usepackage{times}
\usepackage{amssymb,amsbsy,amsmath,amsfonts}
\usepackage{graphicx}
\usepackage{float}
\usepackage{caption}
\usepackage{wrapfig}
\usepackage{subcaption}
\usepackage{color}
\usepackage{morefloats}
\usepackage{rotating}
\usepackage{srcltx}
\usepackage{slashed}
\usepackage{multirow}
\usepackage{verbatim}
\usepackage{tabularx}
\usepackage{adjustbox}
\usepackage[dvipsnames]{xcolor}
\usepackage{nicefrac}
\usepackage{siunitx}

\usepackage{hyperref}
\hypersetup{
	colorlinks	=true,
	urlcolor	=blue,
	linkcolor	=blue,
	citecolor	=blue,
	pdftitle	={},
	pdfauthor	={Zheng-Ting Lai, Jun-Xu Lu, Li-Sheng Geng},
	pdfsubject	={the Fifth Force}
	}

\begin{document}

\title{
Potential of constraining the Fifth Force Using the Earth as a Spin and Mass Source from space
}

\author{Zheng-Ting Lai}
\affiliation{School of Physics, Beihang University, Beijing 102206, China}

\author{Jun-Xu Lu}
\email[Corresponding author: ]{ljxwohool@buaa.edu.cn}
\affiliation{School of Physics, Beihang University, Beijing 102206, China}

\author{Li-Sheng Geng}
\email[Corresponding author: ]{lisheng.geng@buaa.edu.cn}
\affiliation{Sino-French Carbon Neutrality Research Center, \'Ecole Centrale de P\'ekin/School of General Engineering, Beihang University, Beijing 100191, China}
\affiliation{School of Physics, Beihang University, Beijing 102206, China}
\affiliation{Peng Huanwu Collaborative Center for Research and Education, Beihang University, Beijing 100191, China}
\affiliation{Beijing Key Laboratory of Advanced Nuclear Materials and Physics, Beihang University, Beijing 102206, China }
\affiliation{Southern Center for Nuclear-Science Theory (SCNT), Institute of Modern Physics, Chinese Academy of Sciences, Huizhou 516000, China}

\author{Kai Wei}
\affiliation{School of Instrumentation Science and Opto-electronics Engineering, Beihang University, Beijing 100191, China}

\author{Wei Ji}
\affiliation{Johannes Gutenberg University, Mainz 55128, Germany}
\affiliation{Helmholtz Institute Mainz, Mainz 55128, Germany}

\begin{abstract}

We explore the potential of conducting an experiment in a low Earth orbit spacecraft and using the Earth as a spin and mass source to constrain beyond-the-standard-model (BSM) long-range spin- and velocity-dependent interactions, which are mediated by the exchange of an ultralight $\left(m_{Z^{\prime}}<10^{-10}\text{eV}\right)$ or massless intermediate vector boson. The high speed of the low Earth orbit spacecraft can enhance the sensitivity to velocity-dependent interactions. The periodicity enables efficient extraction of signals from background noise, thereby improving the experiment's accuracy. Combining these advantages, we demonstrate theoretically that the novel Spacecraft-Earth model can improve existing bounds on these exotic interactions by up to three orders of magnitude, using the China Space Station (CSS) as a representative low-Earth-orbit carrier. Such a model, if successfully implemented, may provide an innovative strategy for detecting ultralight dark matter and yield tighter constraints on certain coupling constants of exotic interactions.

\end{abstract}


\maketitle


{\it Introduction.}---
The Standard Model (SM) of particle physics has proven to be the most successful theory for describing the fundamental building blocks of nature and their interactions, and it has been extensively tested across multiple fields of physics. However, new physics beyond the SM is eagerly sought to explain unresolved phenomena such as dark matter~\cite{1}, dark energy~\cite{2}, strong CP violation~\cite{3}, the matter-antimatter asymmetry~\cite{4}, and the hierarchy problem~\cite{5}. Among the numerous attempts to extend the Standard Model, the idea of an additional fundamental interaction—beyond the well-known strong, weak, electromagnetic, and gravitational forces—has emerged as one of the most promising solutions. This hypothetical interaction is often referred to as the fifth force~\cite{6,7}.

Hypothetical fifth forces can generally be classified into two categories: those that depend on spins~~\cite{8,9,10} and those that do not~\cite{11}. The former, also known as exotic spin-dependent interactions, could be interpreted as new types of long-range forces mediated between fermions via novel light or massless particles including spin-0 axion-like particles (ALPs)~\cite{3,12,13}, spin-1 bosons~\cite{9}, dark photons~\cite{14,15}, and pairs of fermions~\cite{16,17,18}, all of which are also leading candidates for dark matter or dark energy.  

Moody and Wilczek first proposed the formalism for exotic spin-dependent interactions associated with the axion~\cite{8}. Later, it was shown that any interaction mediated by scalar and vector bosons can be expressed as $16$ terms, each characterized by its spin, mass, distance, and relative velocity~\cite{9}. These potentials were revisited recently~\cite{7,10}, where they were categorized by the types of physical couplings instead of the spin-momentum structure~\cite{9}. Meanwhile, various experiments have been proposed or performed to search for them, such as torsion balance or pendulum experiments~\cite{19,20,21,22,23}, torsional oscillator experiments~\cite{24}, spectroscopy~\cite{25,26}, comagnetometers~\cite{27,28,29,30,31,32}, trapped ions~\cite{33}, nitrogen-vacancy center in diamond~\cite{34,35}, neutron or polarized $^3$He atoms~\cite{36} or other macroscopic experiments~\cite{37,38,39}. These experiments typically utilize a large collection of particles as a polarized spin source or an unpolarized mass source, along with spin sensors — sensitive systems for measuring the resulting shifts in spin energy levels.

Depending on the size of the spin source, these experiments can be roughly divided into two categories: those using laboratory sources and those using the Earth~\cite{31,37,38,39,40,41}, the Moon or the Sun~\cite{42,43,44}, or asteroids~\cite{45}) as an enormous source.  B.R. Heckel et al. utilized the unpolarized matter in the Earth or the Sun to search for exotic velocity-dependent potentials between polarized electrons and unpolarized matter in the Sun and Moon~\cite{42}. Later, L. Hunter et al. utilized the Earth as an enormous polarized geoelectron source to investigate exotic spin- and velocity-dependent interactions~\cite{38,39}. Despite the additional loss due to the distance from the polarized electrons and nucleons to the sensor, the enormous geoelectron and nucleon source significantly improves the sensitivity of detecting exotic spin- and velocity-dependent interactions ~\cite{37,38,39}. Relevant studies include experiments conducted by researchers at the National Institute of Standards and Technology~\cite{46}, the University of Washington~\cite{40}, Amherst College~\cite{37,38,39,47}, Tsinghua University~\cite{48}, and California State University~\cite{41}. We note that the USTC team proposed a similar idea several days after our paper was posted to arXiv~\cite{49}.

Compared to laboratory detection, the main drawbacks of using the Earth as a spin or mass source are the inability to directly modulate the source (though it could be achieved effectively by rotating the experimental apparatus) and artificially control the experiment's location, as well as the limited relative velocity resulting from the Earth's rotation. We notice that these problems can be uniquely resolved by a low Earth orbit spacecraft, with a typical operating altitude of 100 km to 2000 km, corresponding to an overall velocity of 6900 to 7900 meters per second. Specifically, the typical orbital altitude of space stations such as the China Space Station (CSS) and the International Space Station (ISS) is about 400 km. The corresponding nominal orbital velocity is about 7.67 km/s, which offers a relative speed more than 20 times that of ground-based experiments ~\cite{38,39,42}. 

In addition, the orbit of a low Earth orbit vehicle generally covers most of Earth's surface. Taking the CSS as a typical low Earth orbit vehicle, it orbits the Earth from west to east with an orbital period of approximately 90 minutes, and its orbital plane has an inclination of 42–43 degrees to the Earth's equatorial plane. In this way, the CSS covers the Earth's surface between $42^{\circ}$ north and south latitude, allowing experiments to be conducted in diverse locations, a key advantage as emphasized in Refs.~\cite{37,50}. 
Thus, temporally, with low Earth-orbit spacecrafts circling the Earth and the Earth's self-rotation, one anticipates unique periodic signals from the geomagnetic field. Spatially, measurements can be performed at any position, allowing for the exploration of the most stringent constraints with the same sensitivity. 

In the present work, we propose a theoretical "Spacecraft-Earth" model, i.e., conducting an experiment on a low Earth orbit spacecraft, and explore the extent to which the constraints on the coupling constants of the fifth force can be improved solely by virtue of this new model.

{\it Framework.}--- We focus on velocity-dependent interactions in the present work. To better demonstrate the advantages of a space-based experiment, we use the same framework employed by L. Hunter~\cite{38}. Following Ref.~\cite{51}, the six exotic velocity-dependent spin-spin interactions are:
\begin{align}
V_{6,7} = & -\frac{\hbar}{8 \pi c^{2}}\left(\frac{g_{V}^{1} g_{A}^{2}}{2 M_{1}}+\frac{g_{A}^{1} g_{V}^{2}}{2 M_{2}}\right) \nonumber \\
& \times\left[\left(\hat{\boldsymbol{\sigma}}_{1} \cdot \mathbf{v}\right)\left(\hat{\boldsymbol{\sigma}}_{2} \cdot \hat{\mathbf{r}}\right) \pm\left(\hat{\boldsymbol{\sigma}}_{1} \cdot \hat{\mathbf{r}}\right)\left(\hat{\boldsymbol{\sigma}}_{2} \cdot \mathbf{v}\right)\right] \nonumber \\
& \times\left(1+\frac{r}{\lambda}\right) \frac{e^{-r / \lambda}}{r^{2}},
\label{67}
\end{align}
\begin{align}
V_{8} & = \frac{g_{A}^{1} g_{A}^{2}}{4 \pi c^{2}}\left[\left(\hat{\boldsymbol{\sigma}}_{1} \cdot \mathbf{v}\right)\left(\hat{\boldsymbol{\sigma}}_{2} \cdot \mathbf{v}\right)\right] \frac{e^{-r / \lambda}}{r},
\label{8}
\end{align}
\begin{align}
V_{14} & = \frac{g_{A}^{1} g_{A}^{2}}{4 \pi c}\left[\left(\hat{\boldsymbol{\sigma}}_{1} \times \hat{\boldsymbol{\sigma}}_{2}\right) \cdot \mathbf{v}\right] \frac{e^{-r / \lambda}}{r},
\label{14}
\end{align}
\begin{align}
V_{15} = & -\frac{g_{V}^{1} g_{V}^{2} \hbar^{2}}{8 \pi c^{3} M_{1} M_{2}} \nonumber \\
& \times\left[\left(\hat{\boldsymbol{\sigma}}_{1} \cdot(\mathbf{v} \times \hat{\mathbf{r}})\right)\left(\hat{\boldsymbol{\sigma}}_{2} \cdot \hat{\mathbf{r}}\right)+\left(\hat{\boldsymbol{\sigma}}_{1} \cdot \hat{\mathbf{r}}\right)\left(\hat{\boldsymbol{\sigma}}_{2} \cdot(\mathbf{v} \times \hat{\mathbf{r}})\right)\right] \nonumber \\
& \times\left(3+\frac{3 r}{\lambda}+\frac{r^{2}}{\lambda^{2}}\right) \frac{e^{-r / \lambda}}{r^{3}},
\label{15}
\end{align}
\begin{align}
V_{16} = & -\frac{\hbar}{8 \pi c^{3}}\left(\frac{g_{V}^{1} g_{A}^{2}}{2 M_{1}}+\frac{g_{A}^{1} g_{V}^{2}}{2 M_{2}}\right) \nonumber \\
& \times\left[\left(\hat{\boldsymbol{\sigma}}_{1} \cdot(\mathbf{v} \times \hat{\mathbf{r}})\right)\left(\hat{\boldsymbol{\sigma}}_{2} \cdot \mathbf{v}\right)+\left(\hat{\boldsymbol{\sigma}}_{1} \cdot \mathbf{v}\right)\left(\hat{\boldsymbol{\sigma}}_{2} \cdot(\mathbf{v} \times \hat{\mathbf{r}})\right)\right] \nonumber \\
& \times\left(1+\frac{r}{\lambda}\right) \frac{e^{-r / \lambda}}{r^{2}},
\label{16}
\end{align}
where $g_{V/A}$ denotes the vector $(V)$ or axial $(A)$ coupling constant of fermion $1$ or $2$ with spins $\hat{\boldsymbol{\sigma}}_{1/2} $ and masses $M_{1,2}$. The distance and relative speed between the two fermions are denoted by $r$ and $\mathbf{v}$, $\hat{\mathbf{r}}$ is the unit vector of relative coordinate, $\lambda = \hbar / m^{\prime} c$ is the reduced Compton wavelength of the boson of interest of mass $m'$, representing the interaction range of the forces, $\hbar$ is the reduced Planck constant, and $c$ is the speed of light in a vacuum.

\begin{figure}[htbp]
    \centering
    \captionsetup{justification=raggedright,singlelinecheck=false}
    \includegraphics[width=0.35\textwidth]{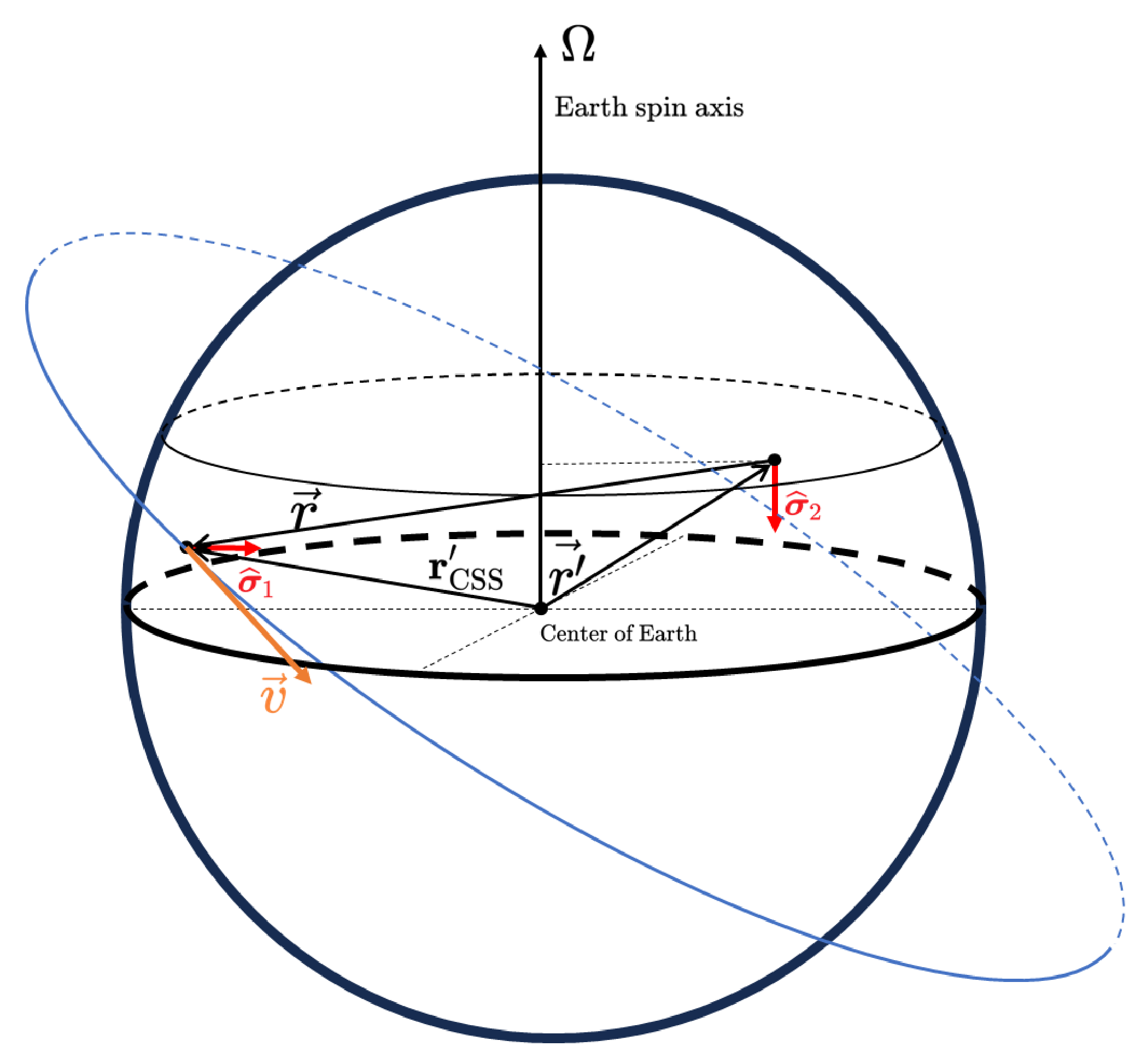}
    \caption{
    Schematic layout of a space-based experiment. $\widehat{\boldsymbol{\sigma}}_{1}$ represents the spin-sensitive direction of the apparatus on a low Earth orbit spacecraft, which is fixed geographically northward(N) or eastward(E), respectively. $\widehat{\boldsymbol{\sigma}}_{2}$ represents the direction opposite to the geomagnetic field. $\mathbf{r}_{\mathrm{s}}^{\prime}$ and $\mathbf{r}^{\prime}$ represent the radial position of the spacecraft and the geoelectron, respectively. $\mathbf{r}_{\mathrm{s}}^{\prime} - \mathbf{r}^{\prime}$ is the relative position.}
    \label{EarthandCSS}
\end{figure}

\begin{figure*}[htbp]
  \centering
  \captionsetup{justification=raggedright,singlelinecheck=false}

  \begin{subfigure}{0.27\textwidth}
    \centering
    \includegraphics[width=\linewidth]{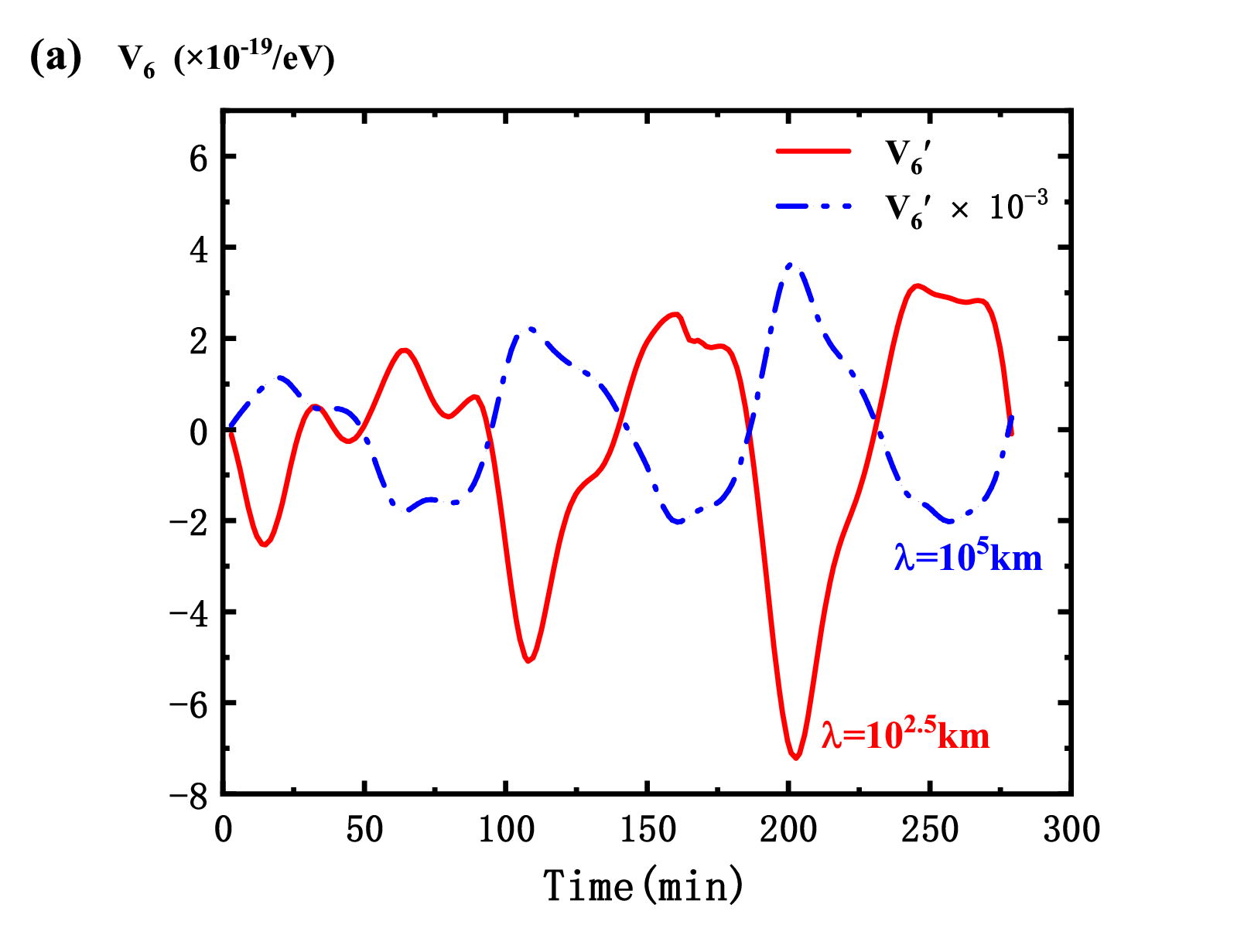}
  \end{subfigure}\hspace{-6mm}
  \begin{subfigure}{0.27\textwidth}
    \centering
    \includegraphics[width=\linewidth]{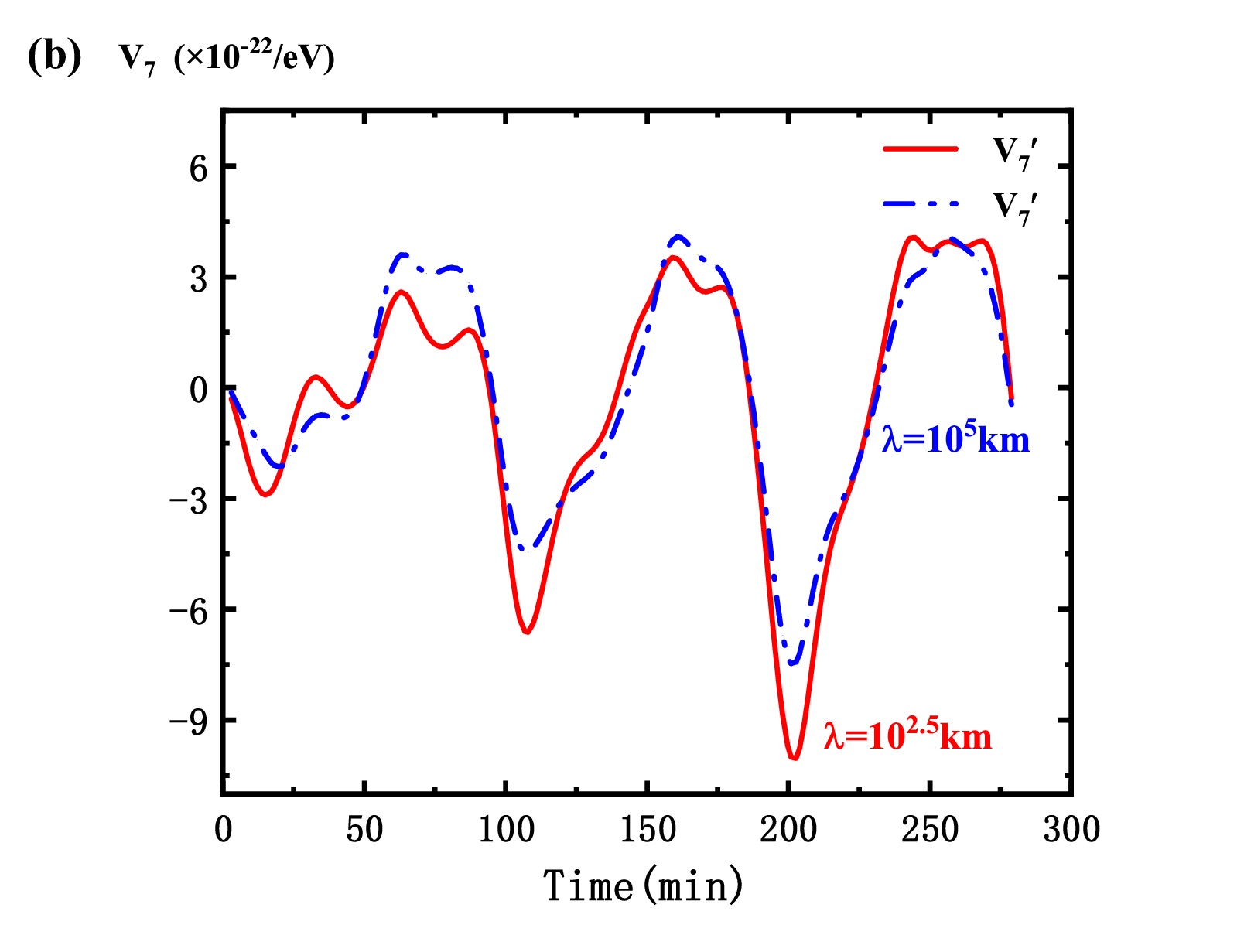}
  \end{subfigure}\hspace{-6mm}
  \begin{subfigure}{0.27\textwidth}
    \centering
    \includegraphics[width=\linewidth]{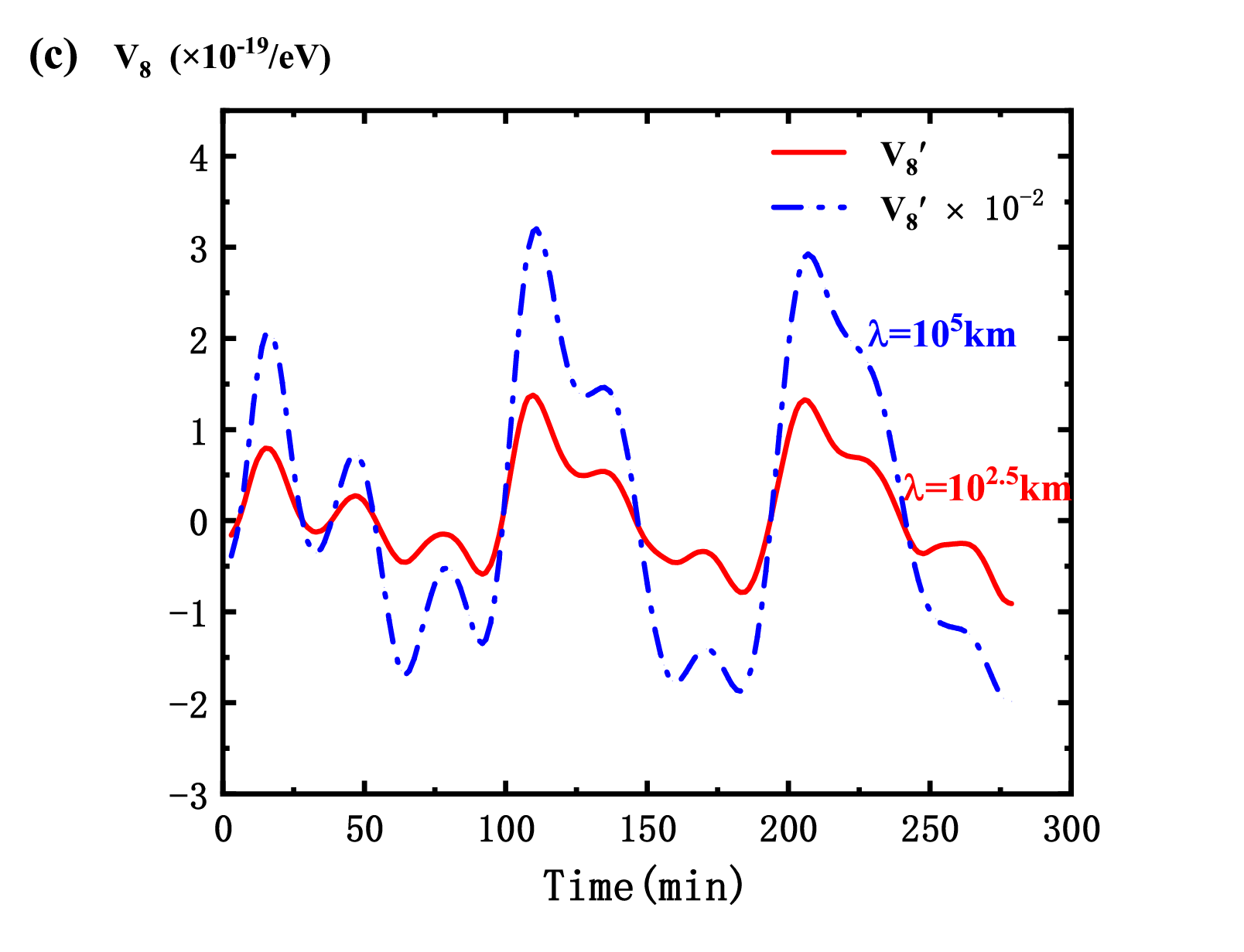}
  \end{subfigure}\hspace{-6mm}
  \begin{subfigure}{0.27\textwidth}
    \centering
    \includegraphics[width=\linewidth]{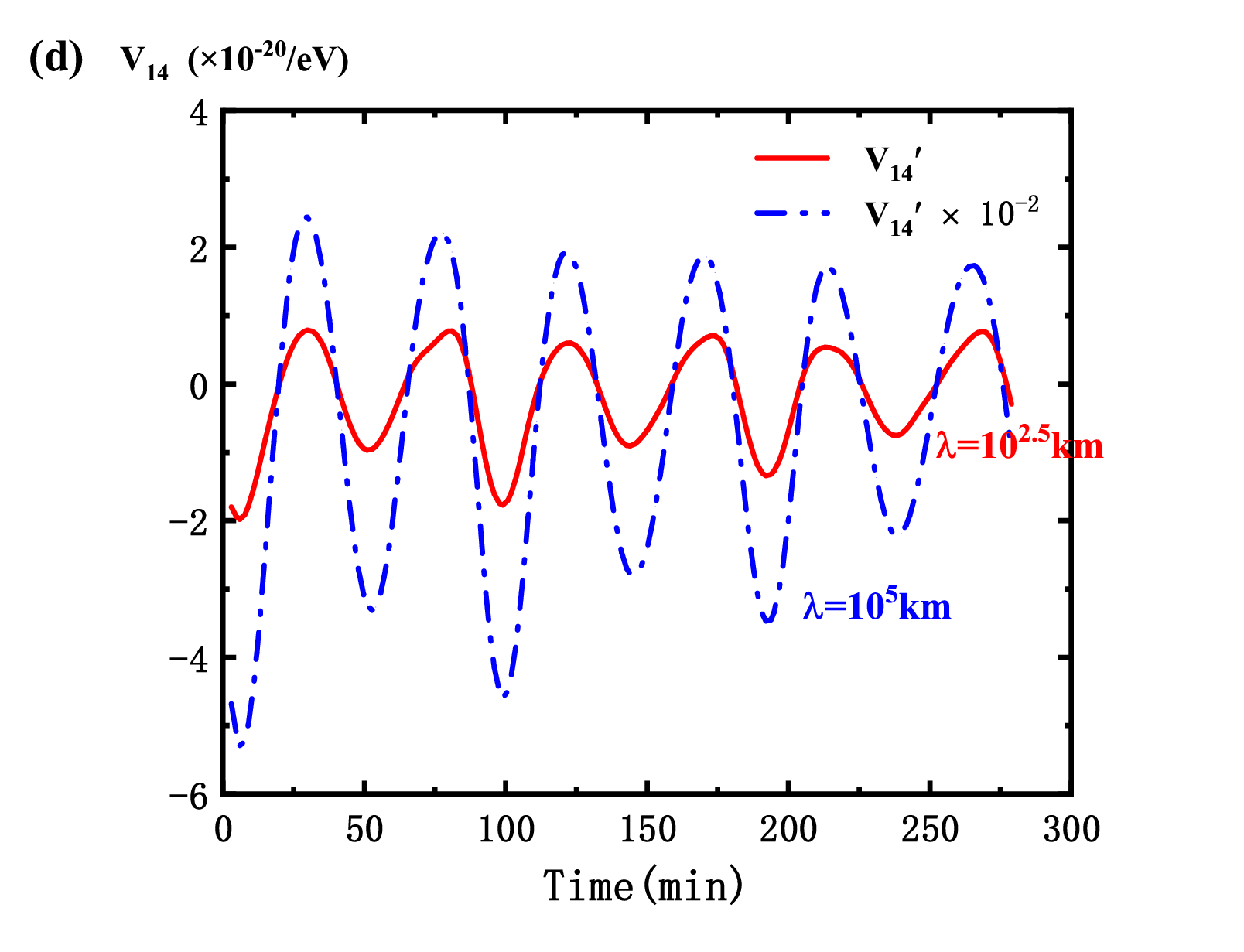}
  \end{subfigure}\vspace{-4mm}

  \begin{subfigure}{0.27\textwidth}
    \centering
    \includegraphics[width=\linewidth]{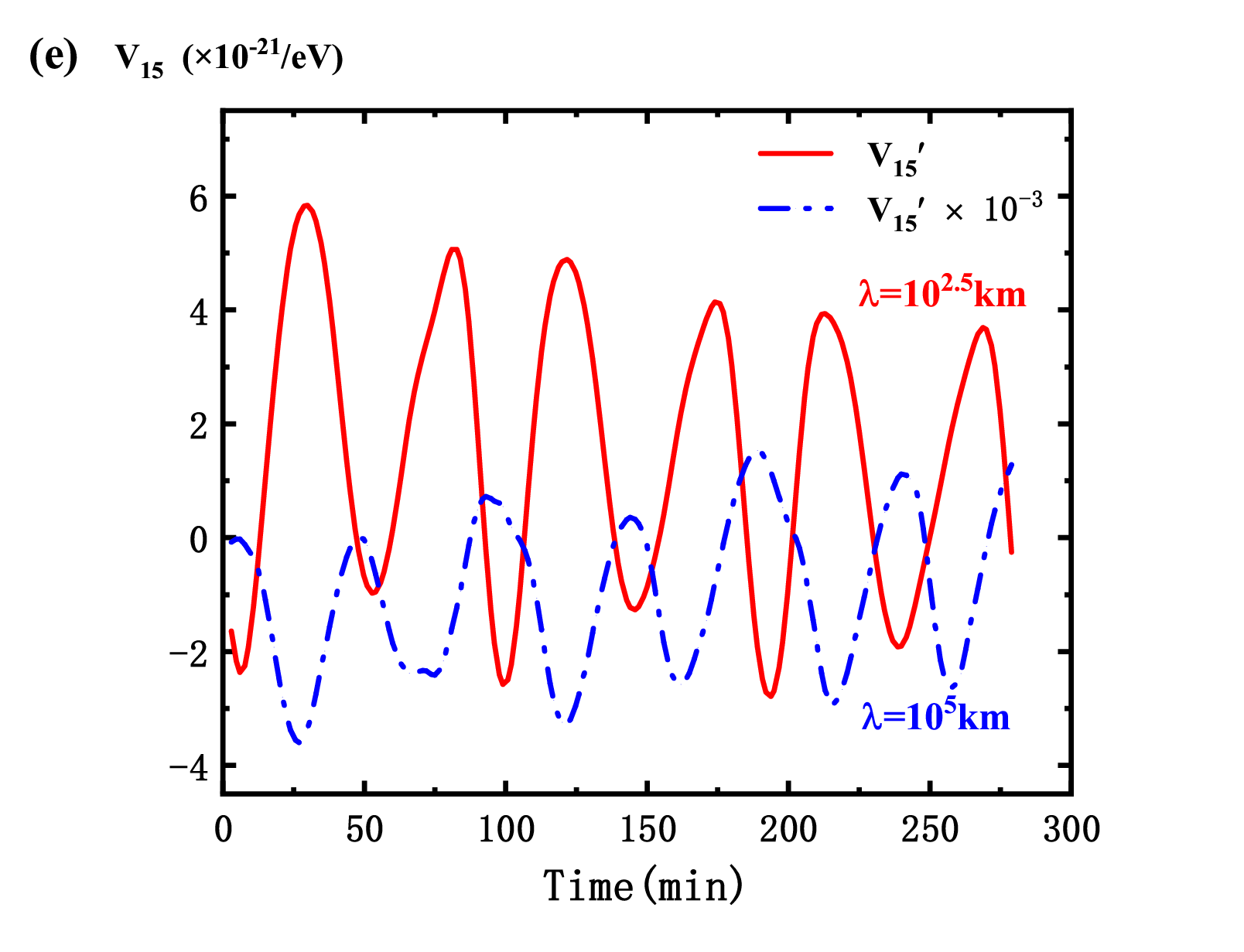}
  \end{subfigure}\hspace{-6mm}
  \begin{subfigure}{0.27\textwidth}
    \centering
    \includegraphics[width=\linewidth]{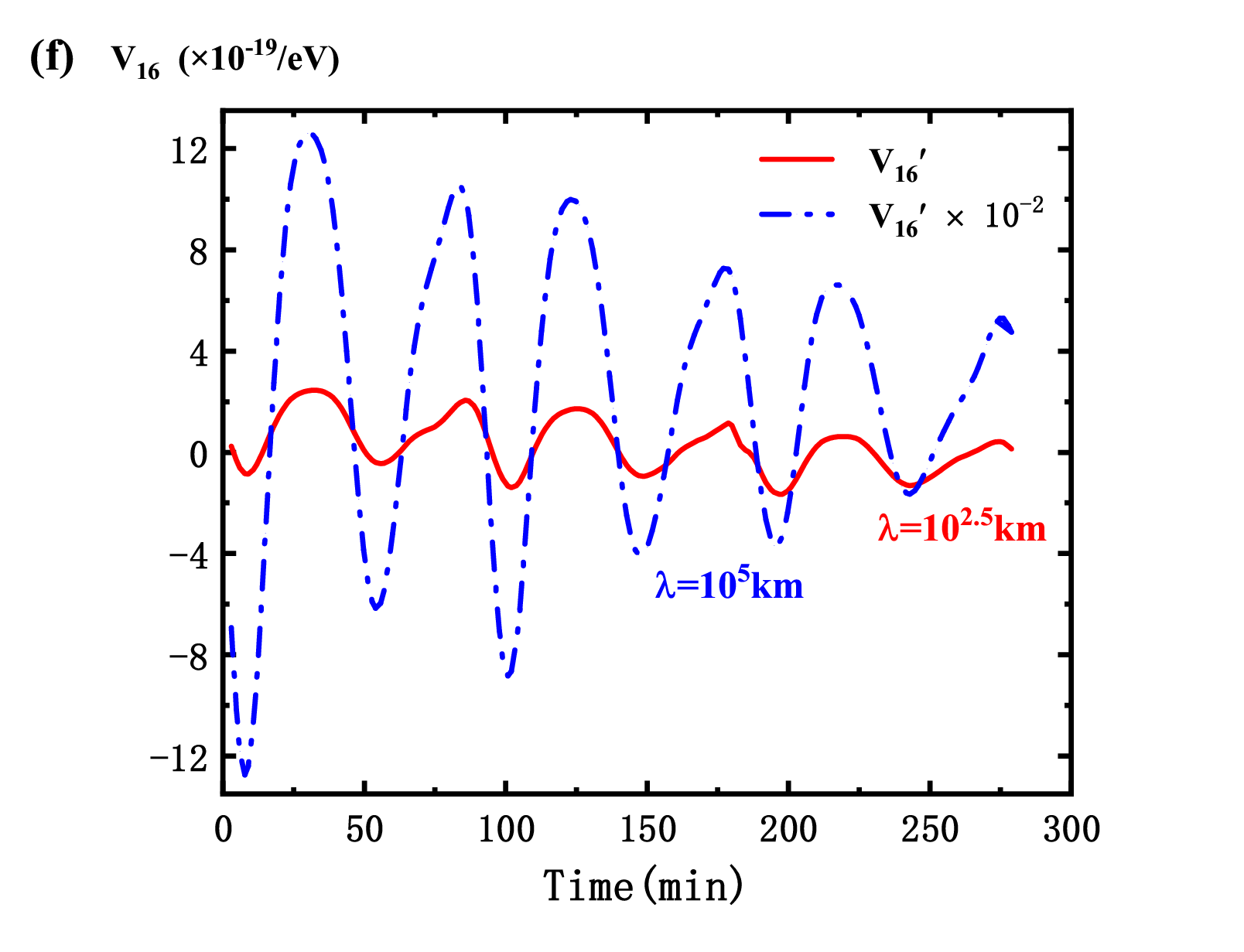}
  \end{subfigure}\hspace{-6mm}
  \begin{subfigure}{0.27\textwidth}
    \centering
    \includegraphics[width=\linewidth]{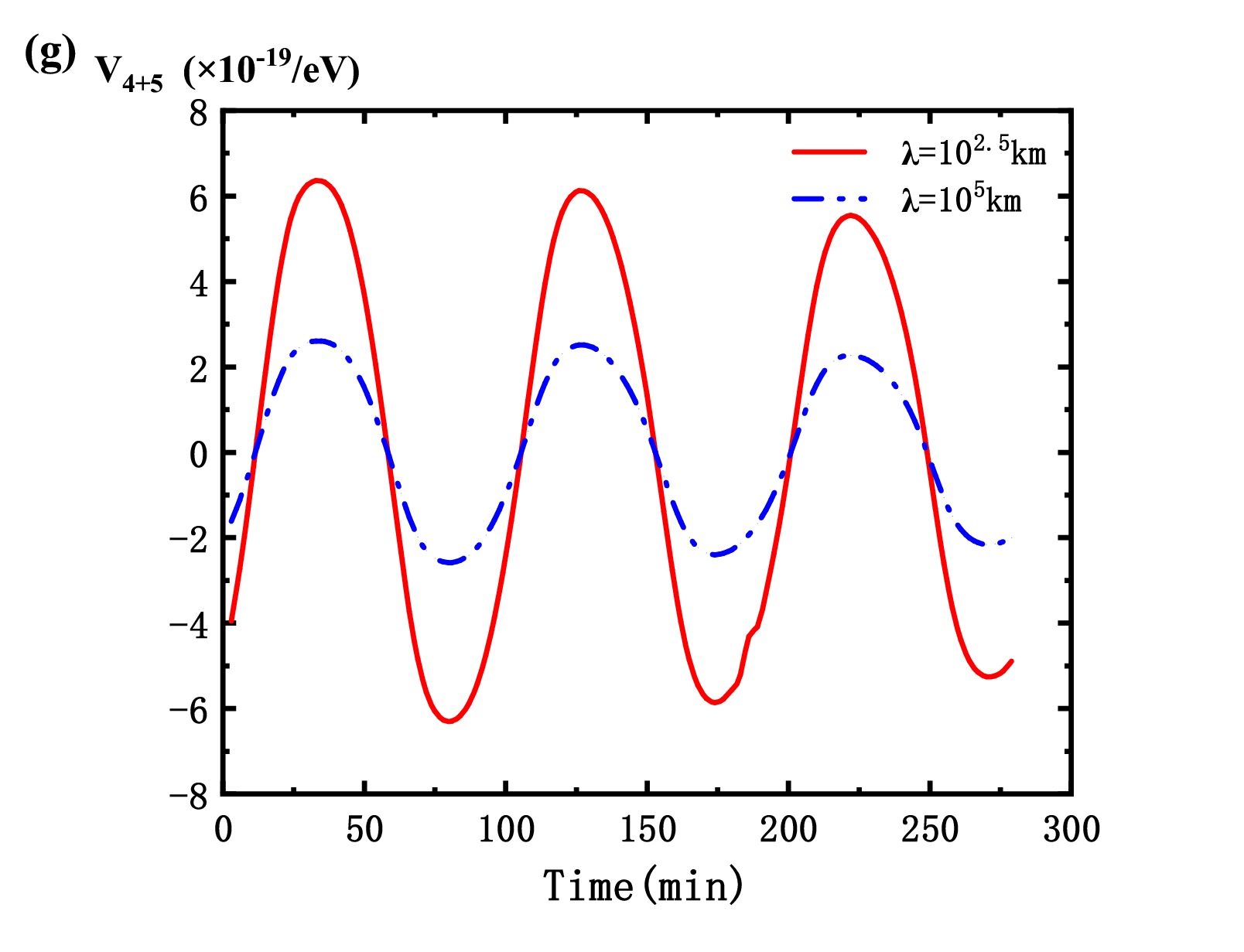}
  \end{subfigure}\hspace{-6mm}
  \begin{subfigure}{0.27\textwidth}
    \centering
    \includegraphics[width=\linewidth]{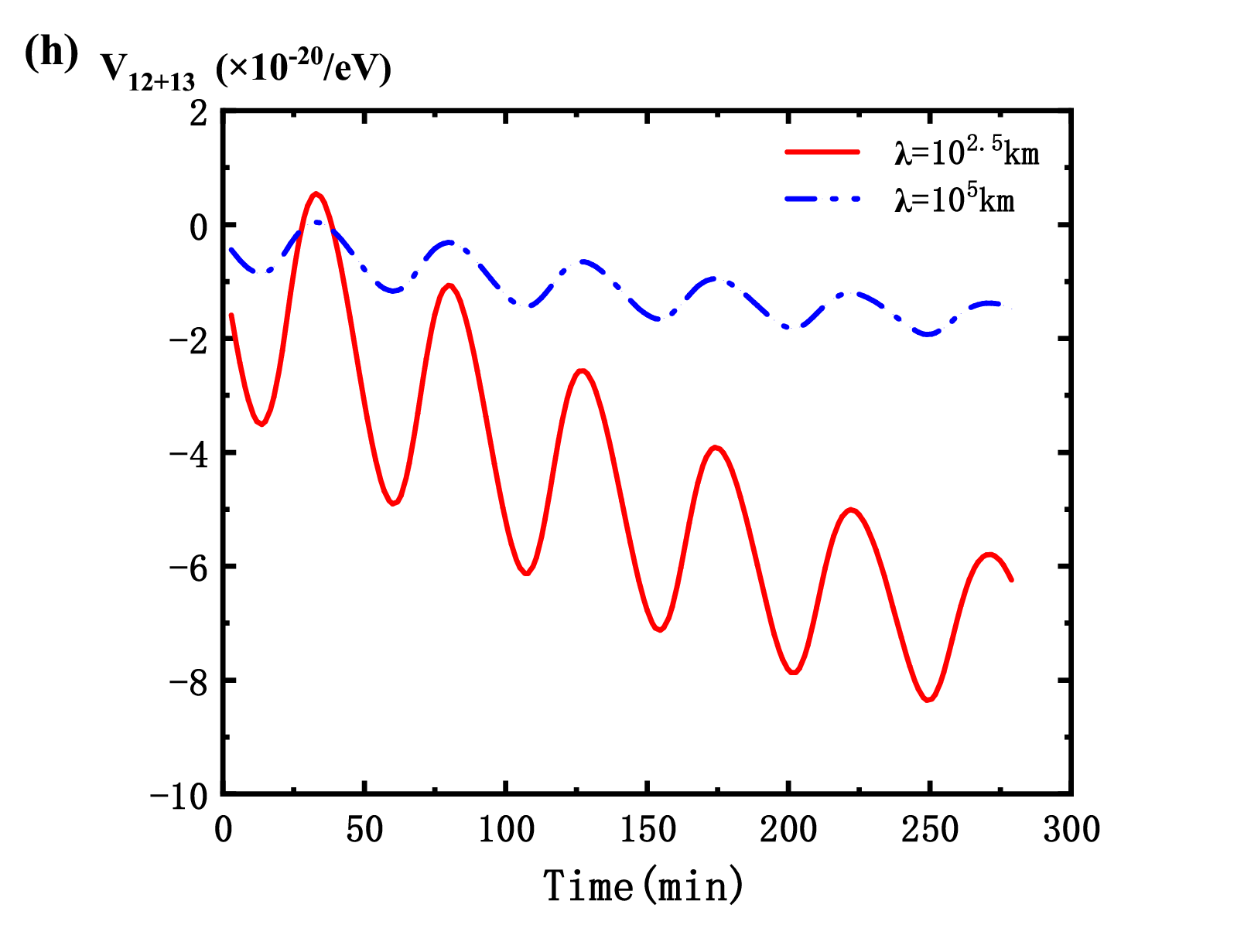}
  \end{subfigure}

  \caption{Periodicity of the expected signals with the assumption that one can use the CSS as the spacecraft. The starting point of the Time axis is arbitrarily chosen to be July $1^{st}$, 2024.}
  \label{Signal}
\end{figure*}

Although $V_{4+5}$ could not provide the most stringent bounds as those of the lowest-order terms, it can probe different spin/velocity structures and confirm which types of couplings the exchanged boson has. Meanwhile, $V_{12+13}$ has the form of a Yukawa potential multiplied by the factor $\boldsymbol{\sigma} \cdot \boldsymbol{v}$, which may provide the only access to some couplings that might vanish in velocity-independent cases~\cite{7}. Similarly, a space-based experiment can also shed light on these two interactions due to the much larger relative velocity, which reads ~\cite{7,39,52}:
\begin{align}
\left.V_{4+5}\right|_{V V} & = \frac{g_{V}^{e} g_{V}^{N}}{16 \pi m_{e}}\frac{\hbar ^{2} }{c}  \boldsymbol{\sigma}_{e} \cdot(\boldsymbol{v} \times \hat{\boldsymbol{r}})\left(\frac{1}{r^{2}}+\frac{1}{\lambda r}\right) e^{-\frac{r}{\lambda}},
\label{4+5VV}
\end{align}
\begin{align}
\left.V_{4+5}\right|_{A A} & = \frac{g_{A}^{e} g_{A}^{N}}{16 \pi}\frac{\hbar ^{2} }{c}  \frac{m_{e}}{m_{N}^{2}} \boldsymbol{\sigma}_{e} \cdot(\boldsymbol{v} \times \hat{\boldsymbol{r}})\left(\frac{1}{r^{2}}+\frac{1}{\lambda r}\right) e^{-\frac{r}{\lambda}},
\label{4+5AA}
\end{align}
\begin{align}
V_{12+13} & = g_{A}^{1} g_{V}^{2} \frac{\hbar}{4 \pi} \boldsymbol{\sigma}_{1} \cdot \boldsymbol{v} \frac{e^{-r / \lambda}}{r}.
\label{12+13}
\end{align}

One should note that velocity-independent interactions can also constrain the coupling constants in Eqs.~(1-8). For example, the $g_Ag_V$ term in $V_{6,7}$, $V_{16}$, and $V_{12+13}$ could also be constrained by $V_{11}$ (following the definition of Ref.~\cite{7}. Actually, the most stringent limit on $g_Ag_V$ for a characteristic length $\lambda > 9.9\times10^{-3}$ m was established by B. R. Heckel et al.~\cite{20} by the $V_{11}$ term and for the range $10^3 < \lambda <10^{11}$m by L. Hunter et al.~\cite{37}. Similarly, the $g_Ag_A$ term in $V_8, V_{14}, V_{4+5}|_{AA}$ could also be constrained by $V_2$ and $V_3|_{AA}$, the $g_Vg_V$ term in  $V_{4+5}|_{AA}$ and $V_{15}$ are further constrained by $V_1, V_3|_{VV}$. For the most stringent limit for these coupling constants from velocity-independent interactions, we refer to Ref.~\cite{7} for a detailed review. In most cases, the velocity-independent interactions, i.e., $V_1, V_2, V_3,V_{11}$, usually provide the most stringent limits for these couplings. But the velocity-dependent terms give rise to distinct types of interactions. Meanwhile, this also means that the methodology to distinguish these interactions remains more challenging.

{\it The ``Spacecraft-Earth" model.}--- We establish the model for the Earth to describe the exotic velocity-dependent interactions between the Earth as the source and the spin sensor on a low Earth orbit spacecraft. The model involves its geoelectron density~\cite{37,53,54,55} and nucleon density ~\cite{39,53,56,57}, which are denoted by $\rho_{e} \left(r^{\prime}\right)$ and $\rho_{n} \left(r^{\prime}\right)$, respectively, the internal temperatures $T\left(r^{\prime}\right)$, where $r^{\prime}$ represents the distance between the source particle and the center of the Earth, as well as the distribution of the geomagnetic field, $\mathbf{B}$, both inside and outside the Earth. 

The model is based on Ref.~\cite{37}, which assumes that Earth’s paramagnetism is predominantly due to unpaired $d$-shell electrons in the iron ions contained within Earth’s mantle and crust minerals. Using this assumption, the geoelectrons' density was calculated. In the model, the density and temperature distributions are assumed to be spherically symmetric, and Earth's core magnetization is neglected, as predicted by density-functional theory calculations~\cite{38}. Additionally, we adopted the temperature model from Ref.~\cite{58,59}, along with the updated World Magnetic Model $($WMM 2020$)$ \cite{60,61}, which is valid from $2020$ to $2025$ for Earth's magnetic field.

The planetary position and velocity of the low Earth orbit spacecraft can be accurately obtained using the Skyfield package ~\cite{62} with the corresponding TLE code ~\cite{63}. In this work, we use $\mathbf{v}\left(r^{\prime}, \theta^{\prime}, \varphi^{\prime}, r_{\mathrm{s}}^{\prime}, \theta_{\mathrm{s}}^{\prime}, \varphi_{\mathrm{s}}^{\prime} \right)$ to represent the relative velocity between the spin sensor on the spacecraft and a specific geoelectron in Earth's mantle. Here, $r^{\prime}$, $ \theta^{\prime}$, $ \varphi^{\prime}$, $ r_{\mathrm{s}}^{\prime}$, $ \theta_{\mathrm{s}}^{\prime}$, and $ \varphi_{\mathrm{s}}^{\prime}$ denote the spherical coordinates of the geoelectron and the spacecraft, respectively, with the center of the Earth as the origin, as shown in Fig.~\ref{EarthandCSS}. The velocity resulting from Earth's rotation is given by $\mathbf{v}^{\prime} = \mathbf{\Omega} \times \mathbf{r}^{\prime}$. By combining this with $\mathbf{v}_{\mathrm{s}}^{\prime}$, we can determine the relative velocity $\mathbf{v} = \mathbf{v}_{\mathrm{s}}^{\prime} - \mathbf{v}^{\prime}$, which is necessary for our calculations.

When summing up the interactions between each polarized geoelectron inside the Earth and the spin sensor on the spacecraft, the total potential can be calculated as follows:
\begin{widetext}
\begin{align} 
V_{\text{total}} = & \int_{0}^{2 \pi} \int_{0}^{\pi} \int_{R_{\text{CM}}}^{R_{s}} r^{\prime 2} \sin \theta^{\prime} \rho_{e} \left(r^{\prime}\right) \frac{2 \mu_{B} B}{k_{B} T\left(r^{\prime}\right)} \times V\left(\mathbf{r}, \mathbf{v}, r^{\prime}, \theta^{\prime}, \varphi^{\prime}, r_{\mathrm{s}}^{\prime}, \theta_{\mathrm{s}}^{\prime}, \varphi_{\mathrm{s}}^{\prime}\right)  d r^{\prime} d \theta^{\prime} d \varphi^{\prime},
\label{vtotal2sigma}
\end{align} 
\end{widetext}
where $\mu_B$ is the Bohr magneton, $k_{B}$ is the Boltzmann constant, and $T$ is the temperature. The integration spans the entire volume from the core-mantle boundary ($R_{\text{CM}}$) to the surface ($R_{s}$). Note that the relative speed $\mathbf{v}$ is the superposition of Earth's rotation and the speed of the low Earth orbit spacecraft, which is far beyond the velocities achievable in laboratory settings ~\cite{39}. 

For $V_{4+5}|_{VV}$, $V_{4+5}|_{AA}$, and $V_{12+13}$, we need to model the Earth as a nucleon source. The expected potentials between the electron and the nucleon read:
\begin{widetext}
\begin{align} 
V_{\text{total}} = & \int_{0}^{2 \pi} \int_{0}^{\pi} \int_{0}^{R_{s}} r^{\prime 2} \sin \theta^{\prime} \rho_{n} \left(r^{\prime}\right) \times V\left(\mathbf{r}, \mathbf{v}, r^{\prime}, \theta^{\prime}, \varphi^{\prime}, r_{\mathrm{s}}^{\prime}, \theta_{\mathrm{s}}^{\prime}, \varphi_{\mathrm{s}}^{\prime}\right)  d r^{\prime} d \theta^{\prime} d \varphi^{\prime},
\end{align} 
\end{widetext}

The difference from Eq.~\ref{vtotal2sigma} is that the $\rho_{n} \left(r^{\prime}\right)$ here represents the density of Earth’s unpolarized nucleons instead of that of polarized electrons. Additionally, the integration should encompass the entire Earth, i.e., from the center of the Earth to the surface ($R_{s}$). 


{\it Results and discussions.}---
Due to the Yukawa-type $e^{-r / \lambda}$ factor, more geoelectrons and nucleons in the mantle contribute as $\lambda$ increases. 
For $\lambda$ well below the altitude of the low Earth orbit spacecraft, i.e., $\lambda \lesssim 100$km, the improvements due to the high speed will be suppressed since the minimal $r$ is the orbit altitude rather than approximately zero, as in ground-based experiments.

In Ref.~\cite{37}, using the measurements from two local Lorentz invariance (LLI) experiments~\cite{42,64}, the bounds on the coupling constants of different velocity-dependent spin-spin interactions between two electrons for various ranges of $\lambda$ were obtained. In the present work, since we are only interested in improvements originating solely from the ``Spacecraft-Earth" model itself, a meaningful comparison between the ground-based and space-based schemes should be conducted under the same sensitivity conditions. To this end, we assume an arbitrary energy sensitivity, e.g., $10^{-20}$ eV for both the ground-based and space-based schemes to demonstrate the improvement from the ``Spacecraft-Earth" model. Note that the energy sensitivity of the LLI experiment~\cite{42,64} has reached $10^{-21}\sim 10^{-22}$ eV over a decade ago. Nowadays, impressive progress has been made in detection techniques, and sensitivity has been reported to improve by several orders of magnitude. We refer to Ref.~\cite{7} for a recent review. However, considering the challenges of the realistic, high-microgravity environment of a space experiment, we assume such a sensitivity ($10^{-20}$ eV), which is slightly inferior to that of the LLI experiment, to conduct our simulations. It is obvious that only the differences between ground-based and space-based experiments are meaningful, and applying a different sensitivity only results in a global shift in the absolute values of all couplings and thus does not alter the relative differences between schemes. 


To demonstrate the signal's periodicity expected in a space-based experiment, we assume the China Space Station (CSS) as a likely experimental platform and take the bounds in Ref.~\cite{37} for electron-electron ($e$-$e$) interactions between velocity-dependent spin-spin potentials as input to extract $V^{\prime}_{\text{total}}$ to be measured. We use $\lambda=10^{2.5} \mathrm{~km}$ (CSS orbital altitude) and $\lambda=10^{5} \mathrm{~km}$ (far exceeding Earth's radius, where potentials saturate) as representative examples to present the results. The expected results corresponding to $\lambda = 10^{2.5}$ km and $\lambda = 10^{5}$ km are shown in Fig.~\ref{Signal}, in which the sensor is supposed to be heading geographically east during detection. In a future experiment, such an oscillating potential could be obtained from the total pseudomagnetic field $\bf{B}_{psd}$ measured by a magnetometer via $V=-\bf{\mu}\cdot\bf{B}_{psd}$, where $\bf{\mu}$ is the magnetic moment of the sensor nucleus or electron. This is actually what is done in ground-based experiments, e.g., Ref.~\cite{32}. 

All these potentials show clear periodicity as the CSS orbits the Earth, which will be beneficial for extracting signals from background noise, thereby improving the experiment's accuracy. Additionally, the line shapes of various interactions differ significantly, suggesting the possibility of distinguishing their contributions in a space-based experiment. This provides unique opportunities to determine the magnitude of the couplings and identify the types of interactions observed.

\begin{figure*}[htbp]
  \centering
  \captionsetup{justification=raggedright,singlelinecheck=false}
  \begin{subfigure}{0.35\textwidth}
    \centering
    \includegraphics[width=\linewidth]{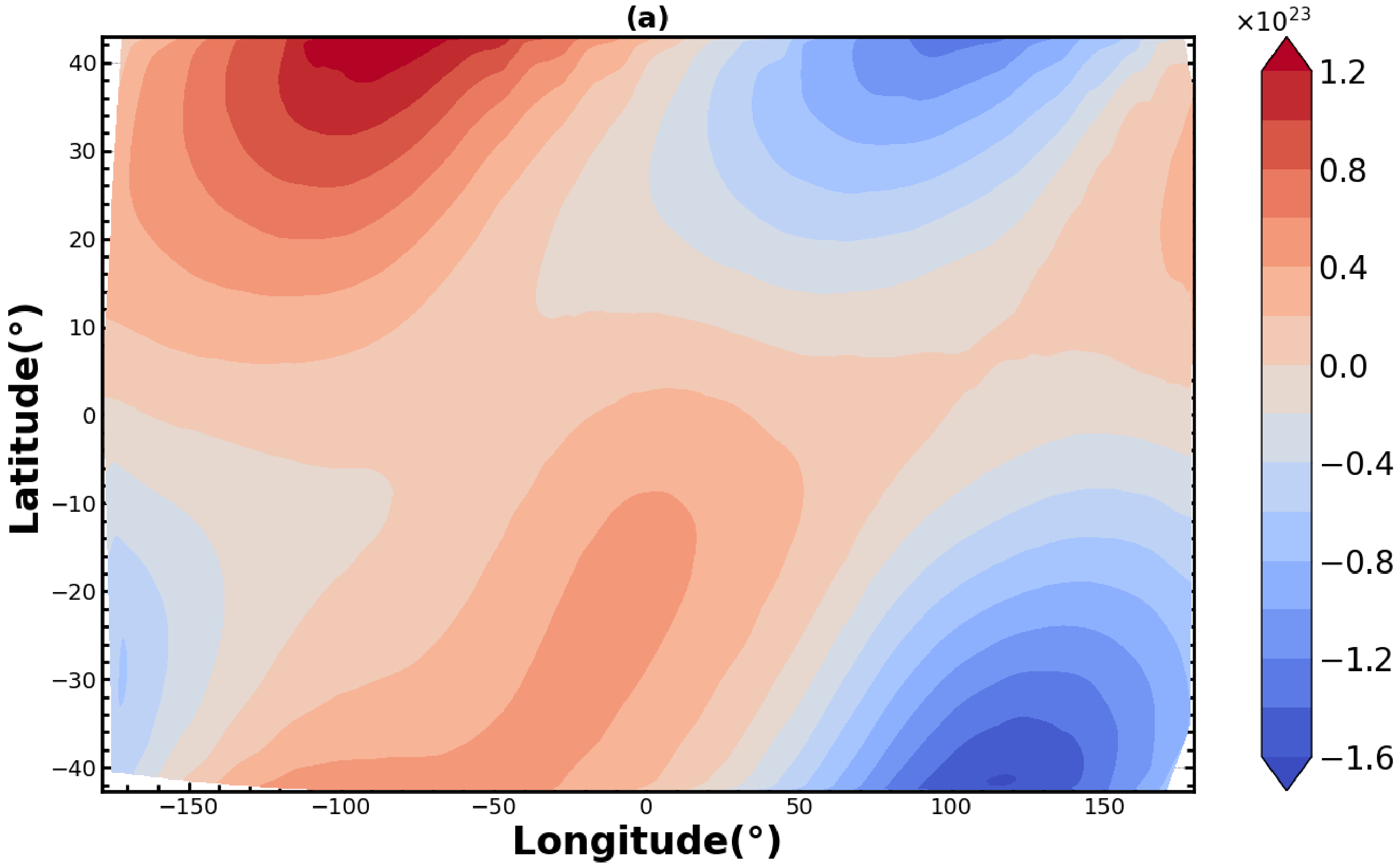}
  \end{subfigure}
  \begin{subfigure}{0.35\textwidth}
    \centering
    \includegraphics[width=\linewidth]{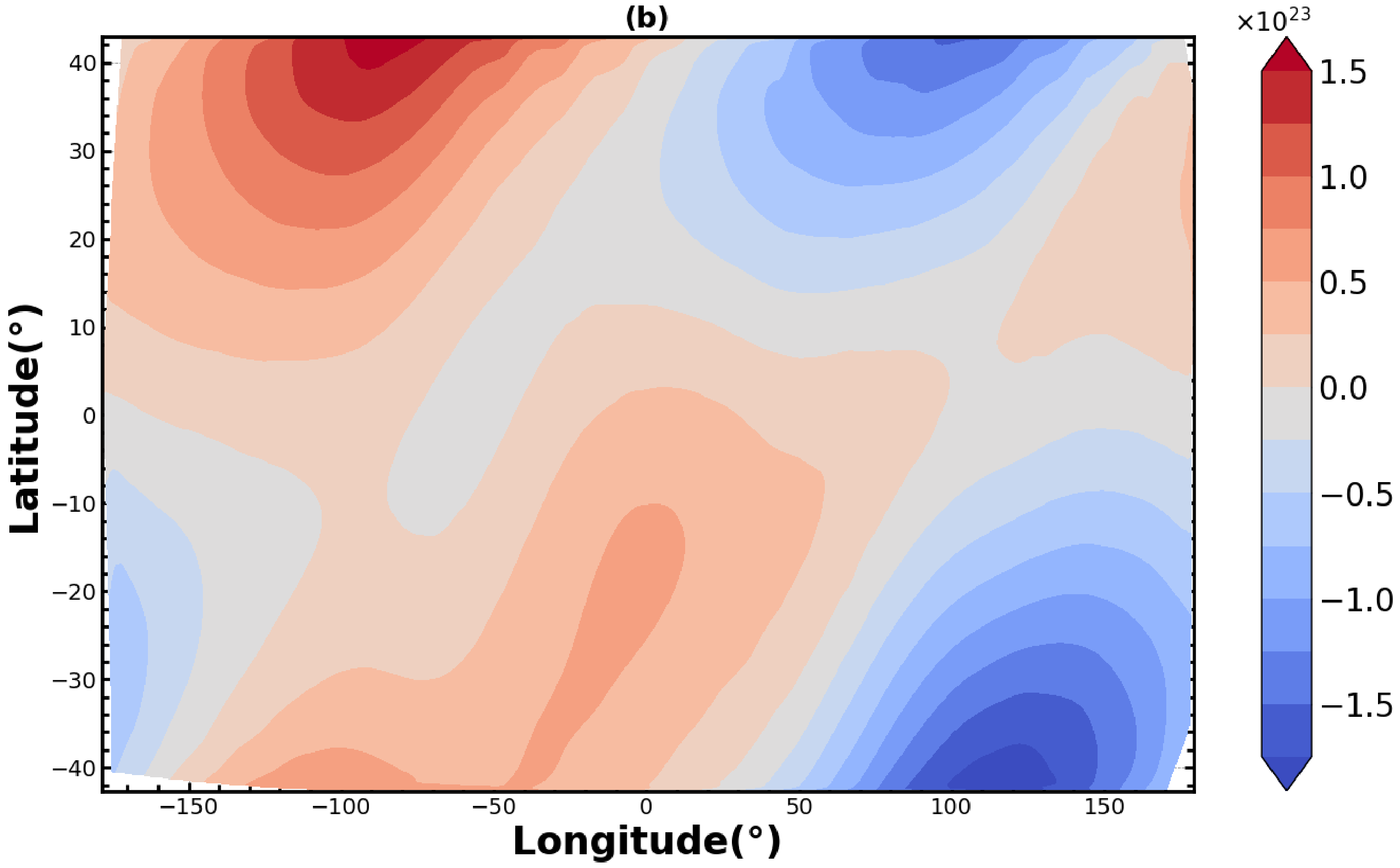}
  \end{subfigure}
  \begin{subfigure}{0.35\textwidth}
    \centering
    \includegraphics[width=\linewidth]{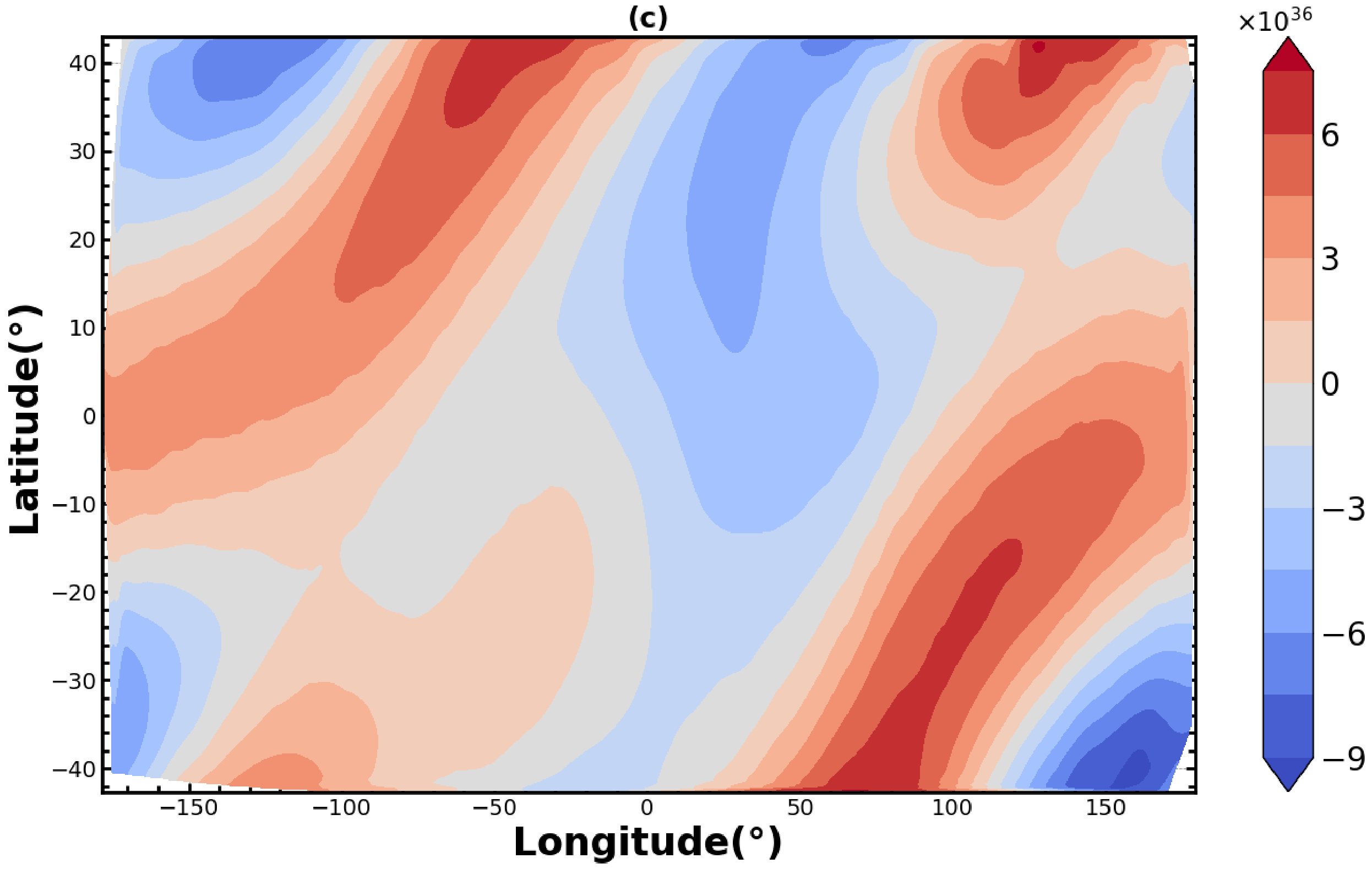}
  \end{subfigure}
  \begin{subfigure}{0.35\textwidth}
    \centering
    \includegraphics[width=\linewidth]{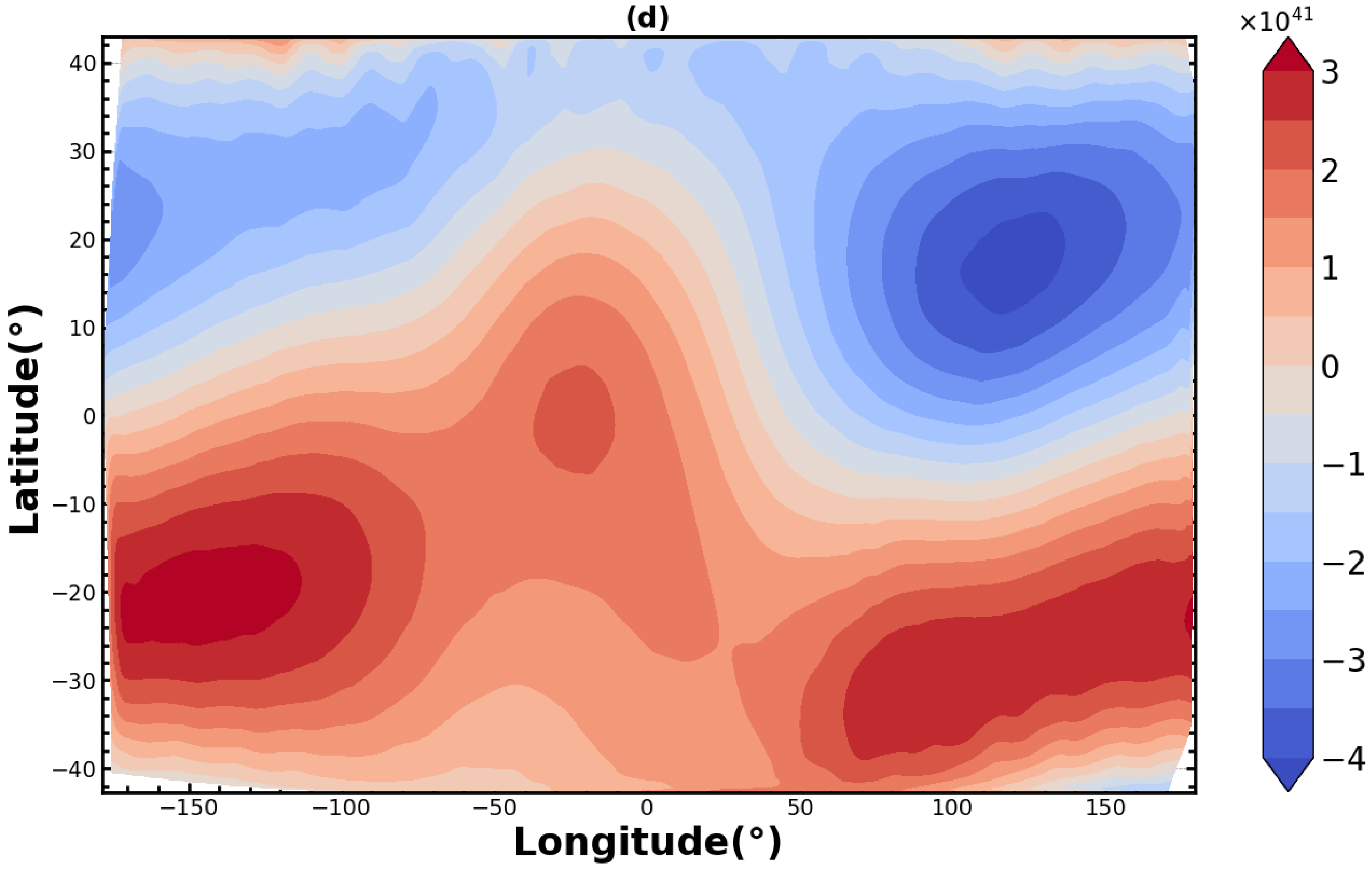}
  \end{subfigure}
  \begin{subfigure}{0.35\textwidth}
    \centering
    \includegraphics[width=\linewidth]{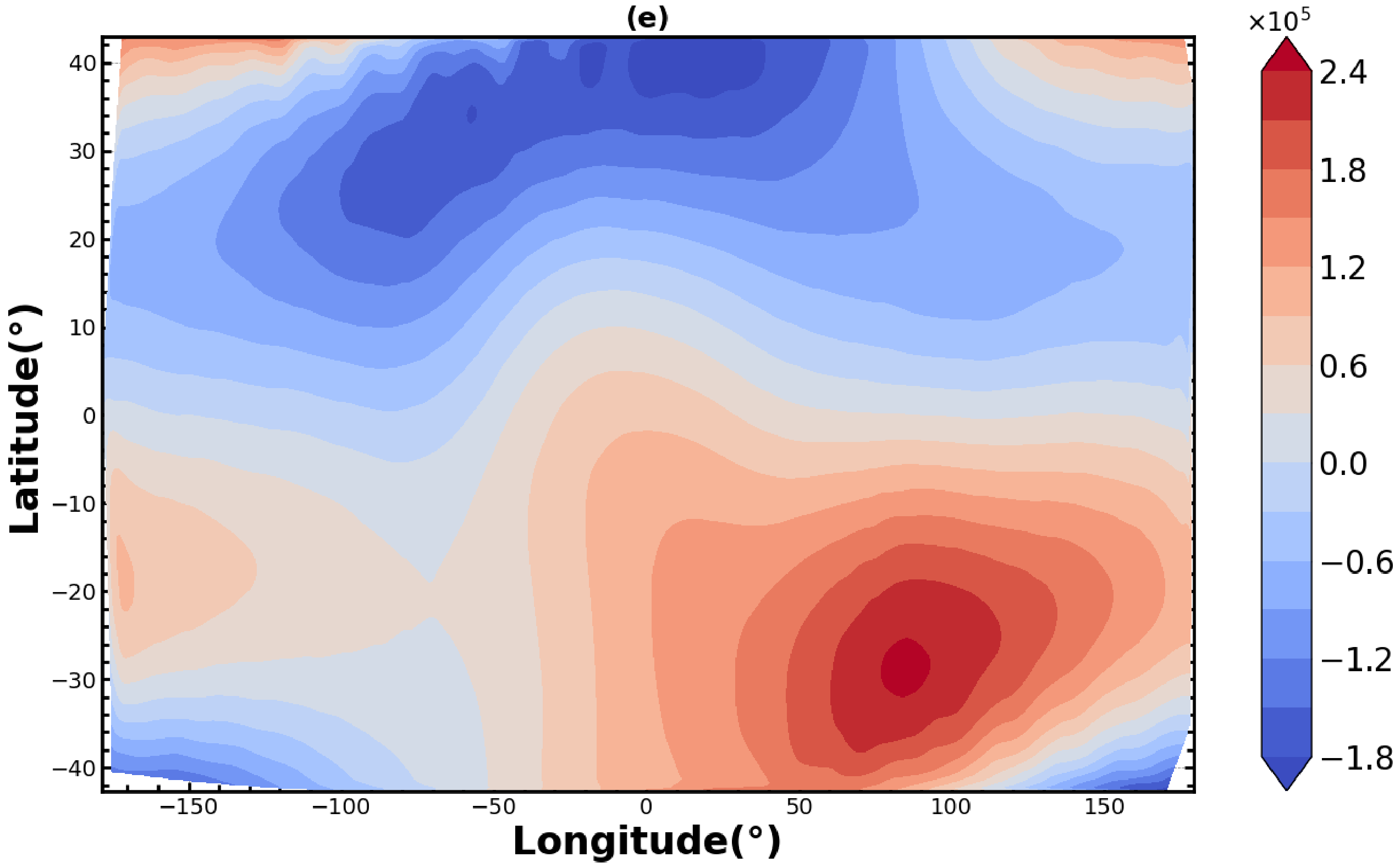}
  \end{subfigure}
  \begin{subfigure}{0.35\textwidth}
    \centering
    \includegraphics[width=\linewidth]{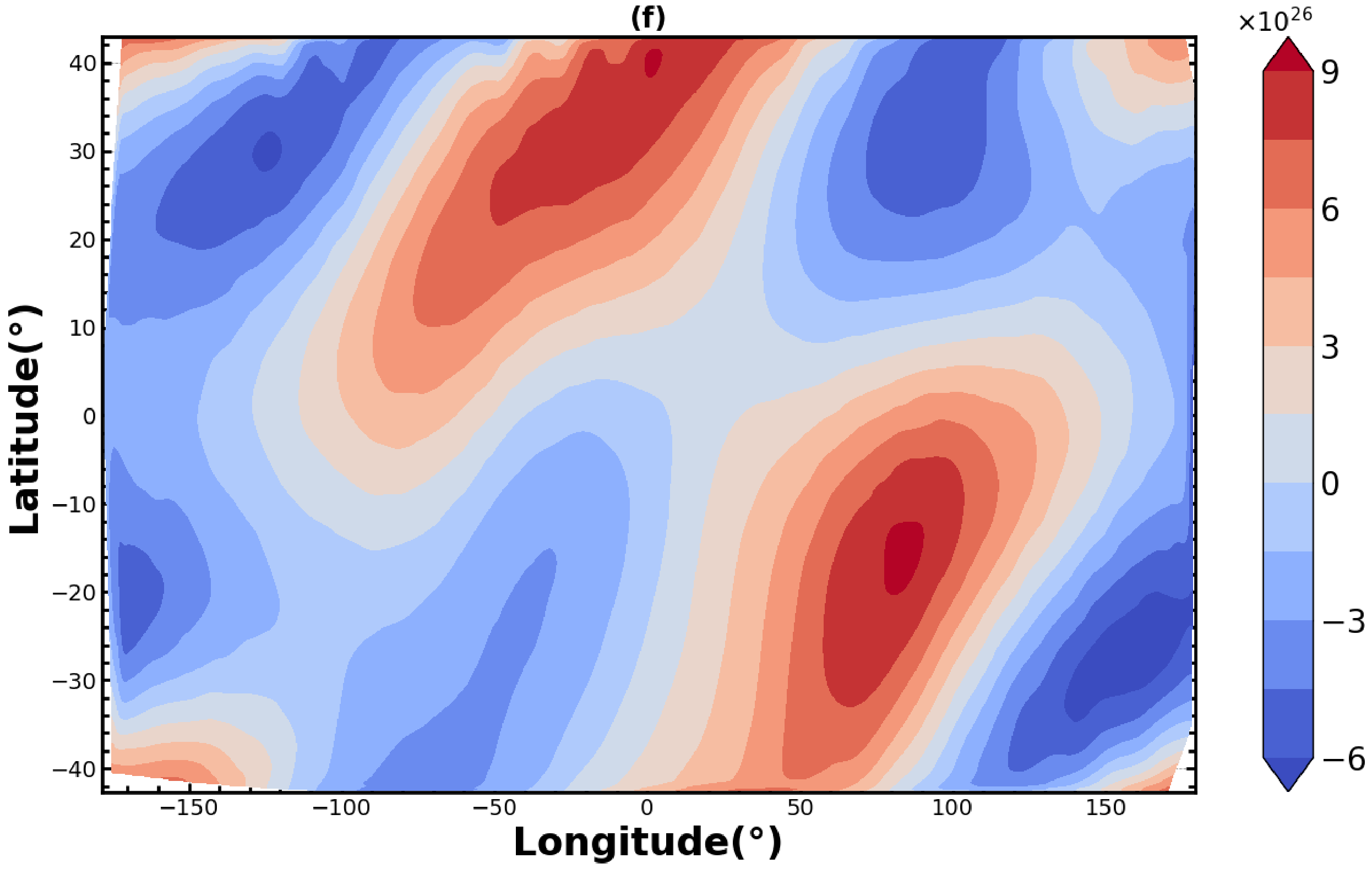}
  \end{subfigure}
  \caption{
  Contour plots of the potentials expected as a spacecraft (CSS) travels from south to north. (a) $V_{6}$. (b)  $V_{7}$. (c)  $V_{8}$. (d)  $V_{14}$. (e)  $V_{15}$. (f)  $V_{16}$.
  }
  \label{Contour}
\end{figure*}

\begin{figure*}[htbp]
  \centering
  \captionsetup{justification=raggedright,singlelinecheck=false}
  \begin{subfigure}{0.365\textwidth}
    \centering
    \includegraphics[width=\linewidth]{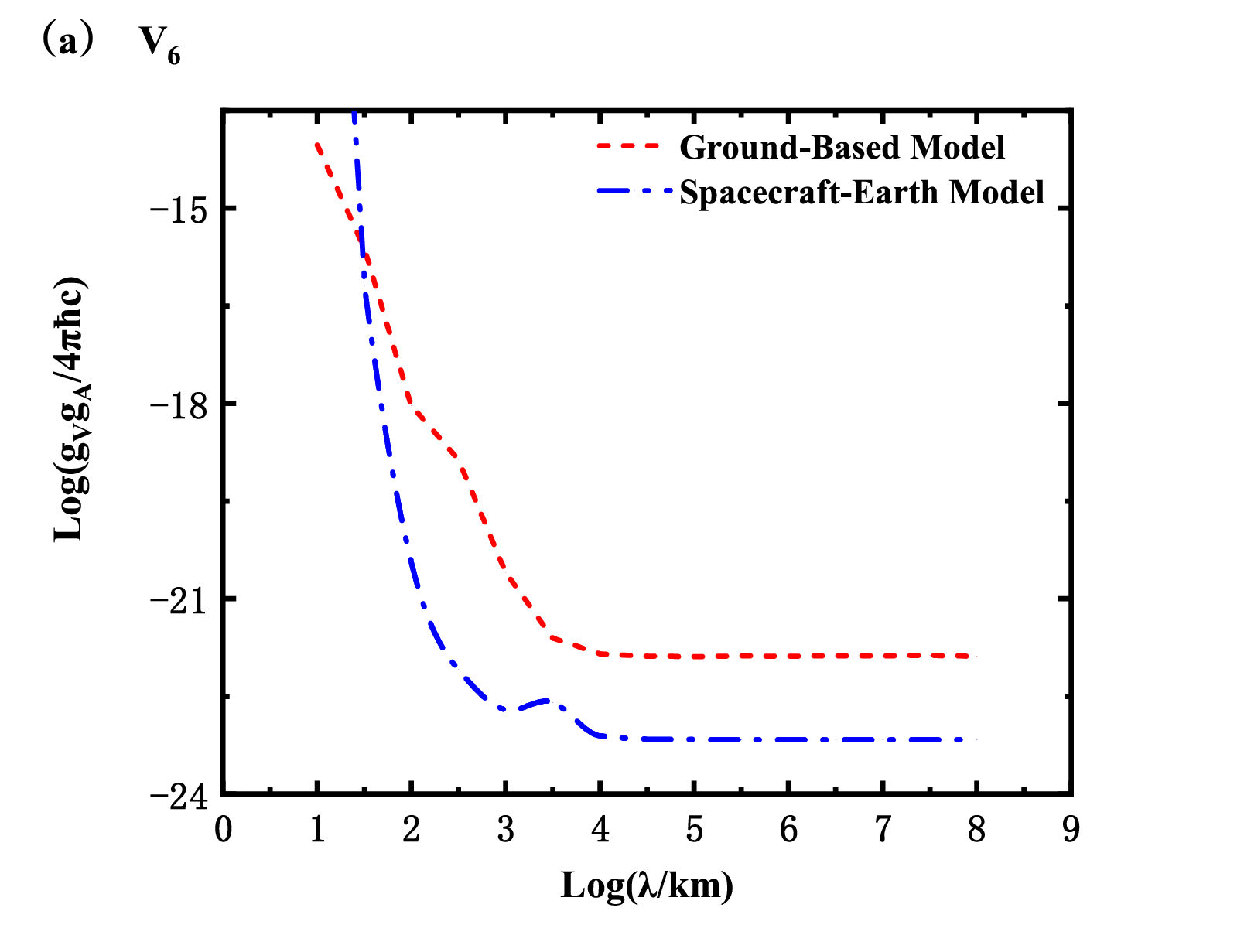}
  \end{subfigure}\hspace{-10mm}
  \begin{subfigure}{0.365\textwidth}
    \centering
    \includegraphics[width=\linewidth]{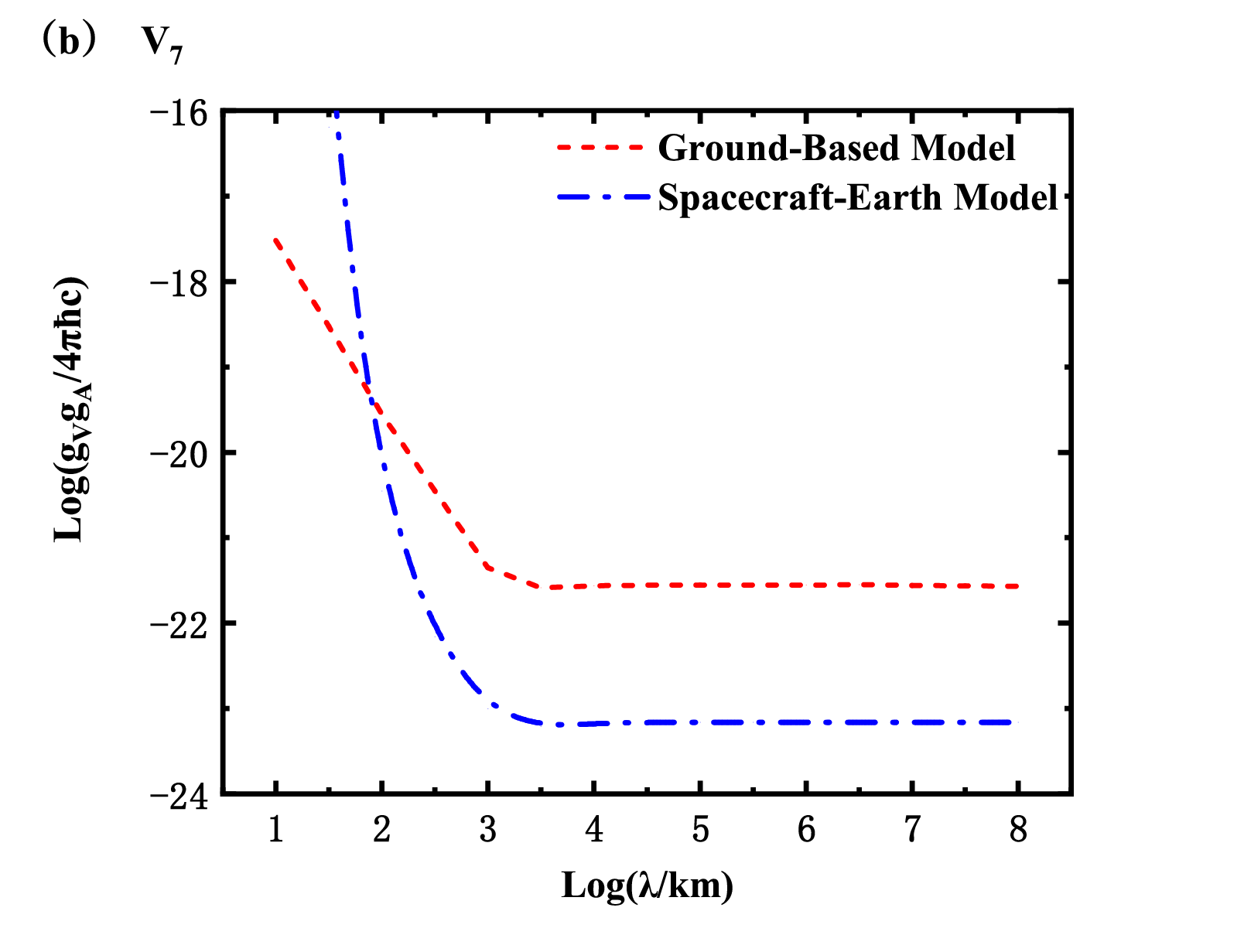}
  \end{subfigure}\hspace{-10mm}
  \begin{subfigure}{0.365\textwidth}
    \centering
    \includegraphics[width=\linewidth]{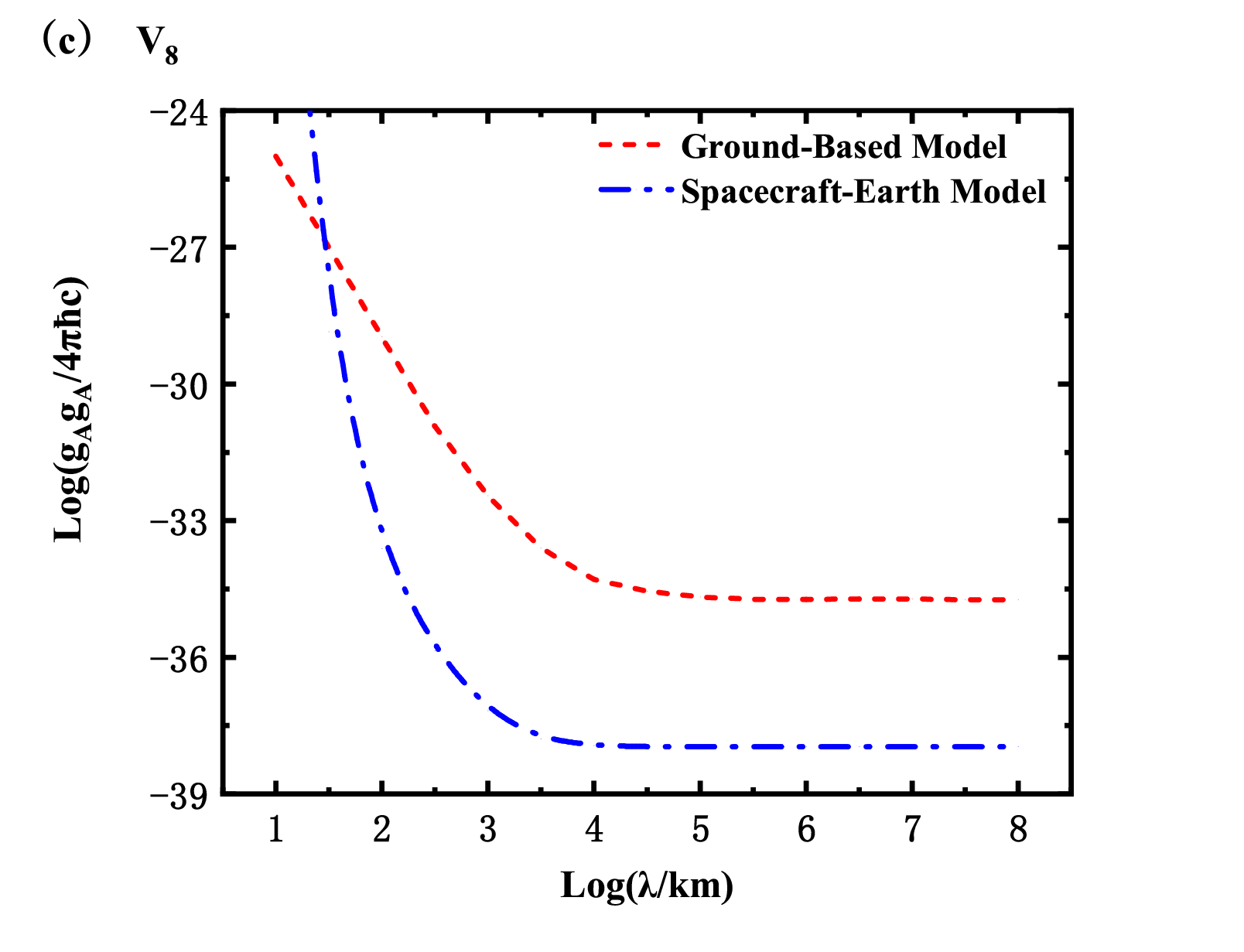}
  \end{subfigure}\vspace{-5mm}

  \begin{subfigure}{0.365\textwidth}
    \centering
    \includegraphics[width=\linewidth]{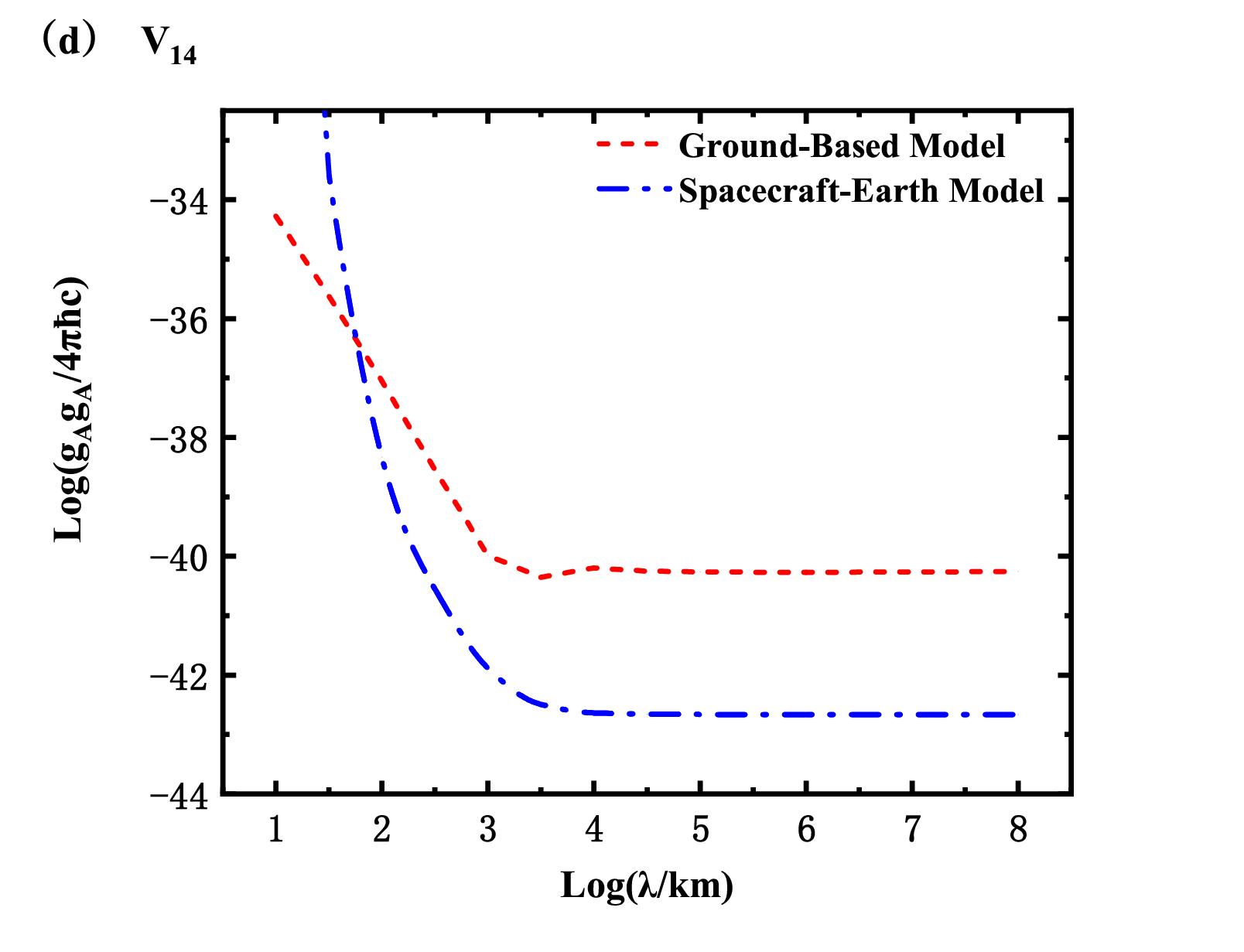}
  \end{subfigure}\hspace{-10mm}
  \begin{subfigure}{0.365\textwidth}
    \centering
    \includegraphics[width=\linewidth]{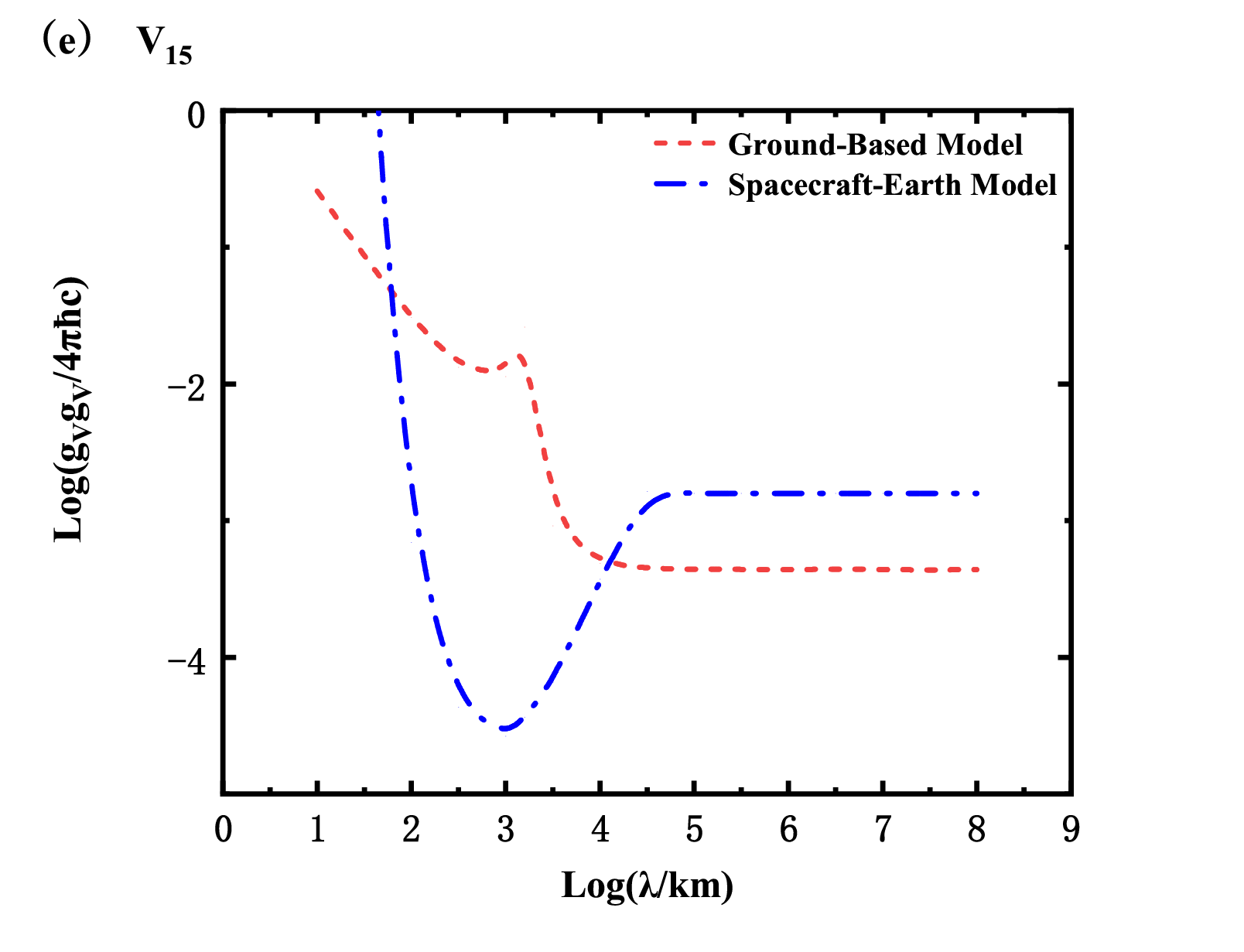}
  \end{subfigure}\hspace{-10mm}
  \begin{subfigure}{0.365\textwidth}
    \centering
    \includegraphics[width=\linewidth]{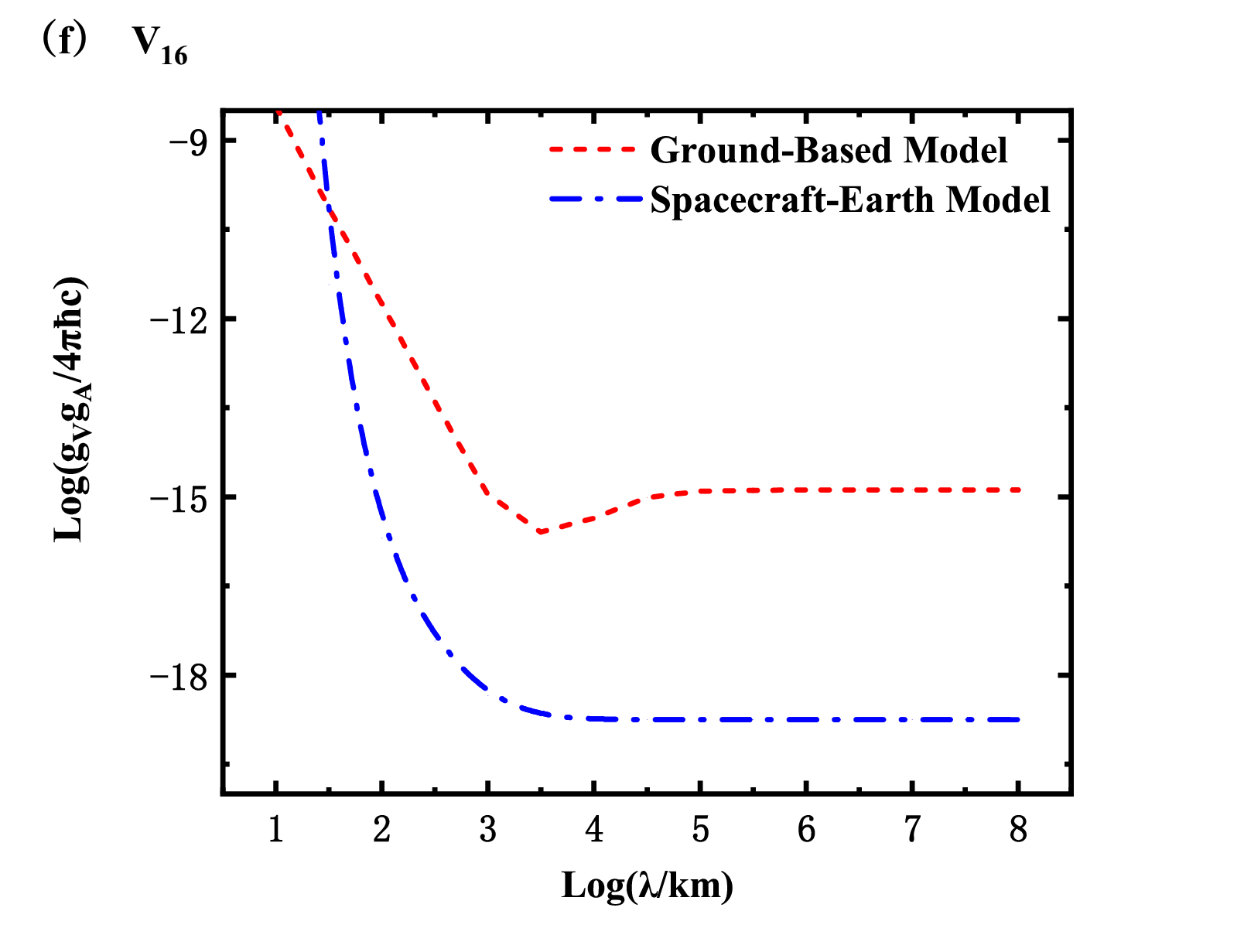}
  \end{subfigure}
  \caption{
  Expected bounds on long-range spin-spin velocity-dependent couplings for electron-electron ($e$-$e$) interactions. The blue curves represent the expected constraints, assuming an arbitrary sensitivity condition (e.g., $10^{-20}$ eV), where the result is the combination of the sensor heading East (E) and North (N) during detection.
  (a) Vector-axial (V-A) couplings [Eq.~\ref{67} with the $+$ sign] for  $V_{6}$. (b)  (V-A)  couplings [Eq.~\ref{67} with the $-$ sign  ] for  $V_{7}$. (c)  (A-A)  couplings [Eq.~\ref{8}] for  $V_{8}$. (d) (A-A) couplings [Eq.~\ref{14}] for  $V_{14}$. (e)  (V-V)  couplings [Eq.~\ref{15}] for  $V_{15}$. (f) (V-A) couplings [Eq.~\ref{16}] for  $V_{16}$.}
  \label{bound}
\end{figure*}

Once the spacecraft's trajectory covers the entire area, a contour plot can be generated using the corresponding Two-Line Element (TLE) code, such as the TLE for the CSS~\cite{63}. And the most sensitive position for each interaction can be addressed. Because of the periodic orbital motion, signal-processing techniques can be applied to enhance detection accuracy.  

Again, we assume that the experiment will be conducted on the CSS and use its TLE code~\cite{63} mentioned earlier to generate the contour plot in Fig.~\ref{Contour}. Due to the vector-based geographic relationships in the calculations, the potential measured at a given position may vary depending on the direction from which the CSS approaches. This means that the potential detected when the CSS travels from south to north may differ from when it travels from north to south. However, as long as the TLE data~\cite{63} remains unchanged, the contour plot will stay fixed. Consequently, as the CSS orbits the Earth, the signals it receives will exhibit periodicity with each orbit. Although slight differences in the signals may arise from variations in angular velocity, the overall periodicity is sufficient to apply signal-processing techniques to distinguish the signals from background noise, thereby improving the experiment's accuracy.

In previous work~\cite {37,50}, it was proposed that conducting experiments in southern Thailand would yield approximately twice the sensitivity compared with those performed in Amherst, owing to the stronger and more parallel surface magnetic field in Thailand. This highlights the importance of the experiment's location. In our study, the flexibility of the low Earth orbit enables it to cover different regions over time, allowing the identification of optimal locations for detecting new interactions at varying velocities. We believe such an experiment will be highly effective in pinpointing the optimal regions for achieving the strongest signals, thereby significantly improving the sensitivity limits. 

Compared to the works by L. Hunter et al. ~\cite{38,37,64}, we found that the bounds on the electron ($e$) energy, when its spin is oriented north (N), are generally less restrictive than when oriented east (E). This suggests that the most restrictive bounds are likely to be obtained through measurements in the eastward direction. Another key difference is that, unlike experiments conducted on Earth, the potential of long-range interactions does not always maintain the same sign in space. As a spacecraft orbits the Earth, the angles between $\widehat{\boldsymbol{\sigma}}_{1}$, $\widehat{\boldsymbol{\sigma}}_{2}$, $\mathbf{r}$, and $\mathbf{v}$ continuously change. As a result, it is natural to observe the potential oscillating between positive and negative values, as illustrated in Fig.~\ref{Signal}.

In Fig.~\ref{bound}, we present the expected uppest bounds for all six exotic velocity-dependent spin-spin interactions' couplings for $e$-$e$ interactions as functions of the effective range $\lambda$, assuming an arbitrary energy resolution, e.g., $10^{-20}$ eV which will not exert any impact on the conclusions.
Similar to the conclusion obtained in the ground-based experiment~\cite{38}, the coupling bounds of $e$-$e$ interaction based on the new model with spin-oriented north (N) are usually not as stringent as those with spin-oriented east (E). The coupling constants for most interactions are expected to be more constrained by at least three orders of magnitude. In those less sensitive regions, an improvement of at least one order of magnitude can still be expected. 


The bounds on the couplings of $V_{6}$ (Fig.~\ref{bound}(a)) and $V_{7}$ (Fig.~\ref{bound}(b)) are lowered significantly. For $V_6$, at $\lambda = 10^{2.5}$ km, the new model results in a bound approximately 2.4 orders of magnitude more restrictive, and in the flatter regions, the improvement is at least 1.3 orders of magnitude. For $V_7$, the coupling bounds are reduced by about 1.6 orders of magnitude across the full parameter range. Furthermore, the "Spacecraft-Earth" model enables the measurement of $V_{6}$ oriented east for $\lambda$ less than $10^{2}$ km, which is inaccessible to previous experiments~\cite{38}. Moreover, since $V_{6+7}$ would be more meaningful than $V_6$ or $V_7$ individually for measurements~\cite{7}, one can further constrain $V_{6+7}$, which reads:
\begin{align}
V_{6+7} &= -\frac{\hbar}{4 \pi c^{2}}\left(\frac{g_{V}^{1} g_{A}^{2}}{2 M_{1}}+\frac{g_{A}^{1} g_{V}^{2}}{2 M_{2}}\right)  \times\left[\left(\hat{\boldsymbol{\sigma}}_{1} \cdot \mathbf{v}\right)\left(\hat{\boldsymbol{\sigma}}_{2} \cdot \hat{\mathbf{r}}\right)\right] \nonumber\\
&\times\left(1+\frac{r}{\lambda}\right) \frac{e^{-r / \lambda}}{r^{2}}
\label{6+7}
\end{align}

\begin{figure}[htbp]
    \centering
    \captionsetup{justification=raggedright,singlelinecheck=false}
    \includegraphics[width=0.45\textwidth]{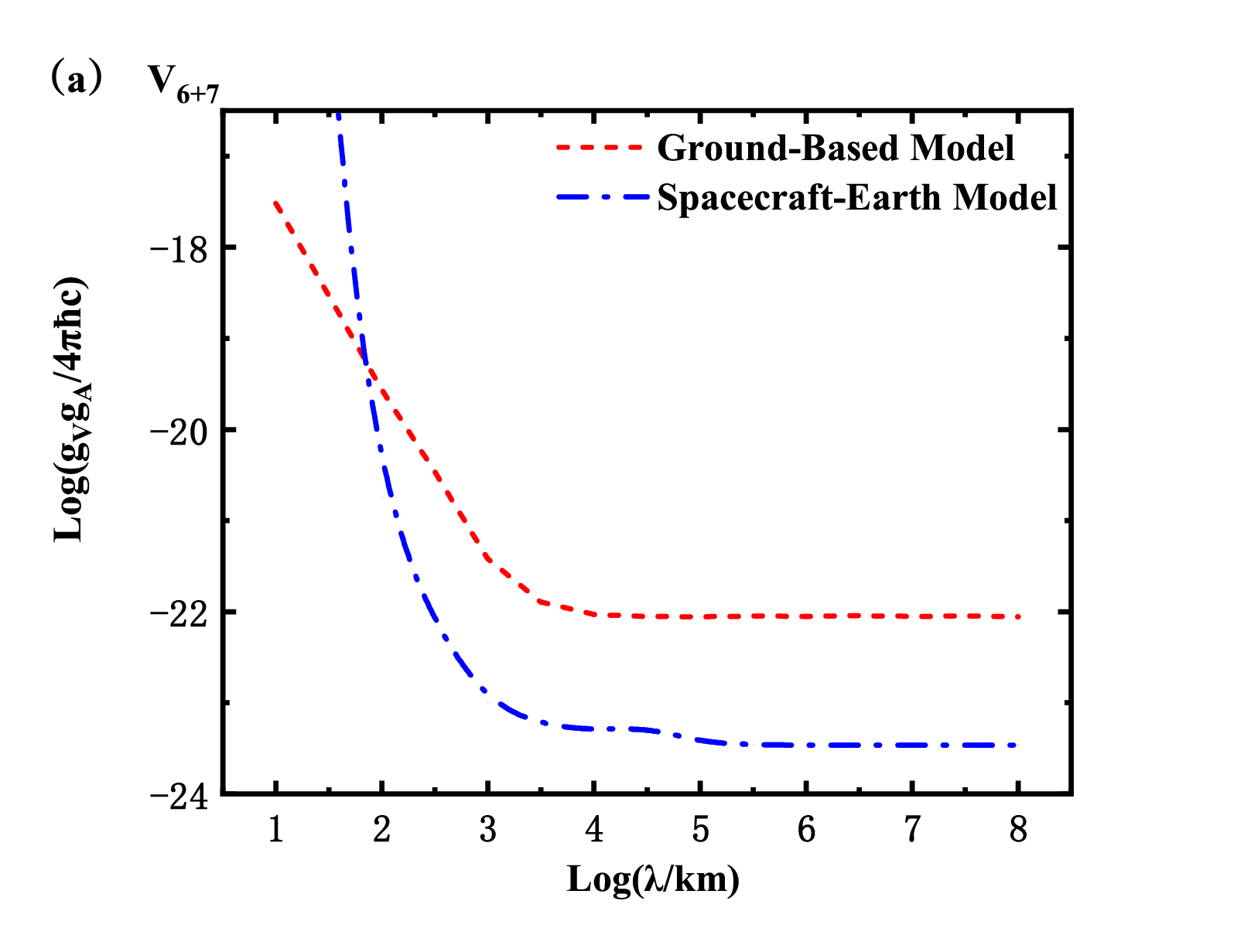}
    \caption{
    Same as Fig.~\ref{bound} but for $V_{6+7}$.
    }
    \label{V67RESULT}
\end{figure}

\begin{figure*}[htbp]
  \centering
  \captionsetup{justification=raggedright,singlelinecheck=false}
  \begin{subfigure}{0.365\textwidth}
    \centering
    \includegraphics[width=\linewidth]{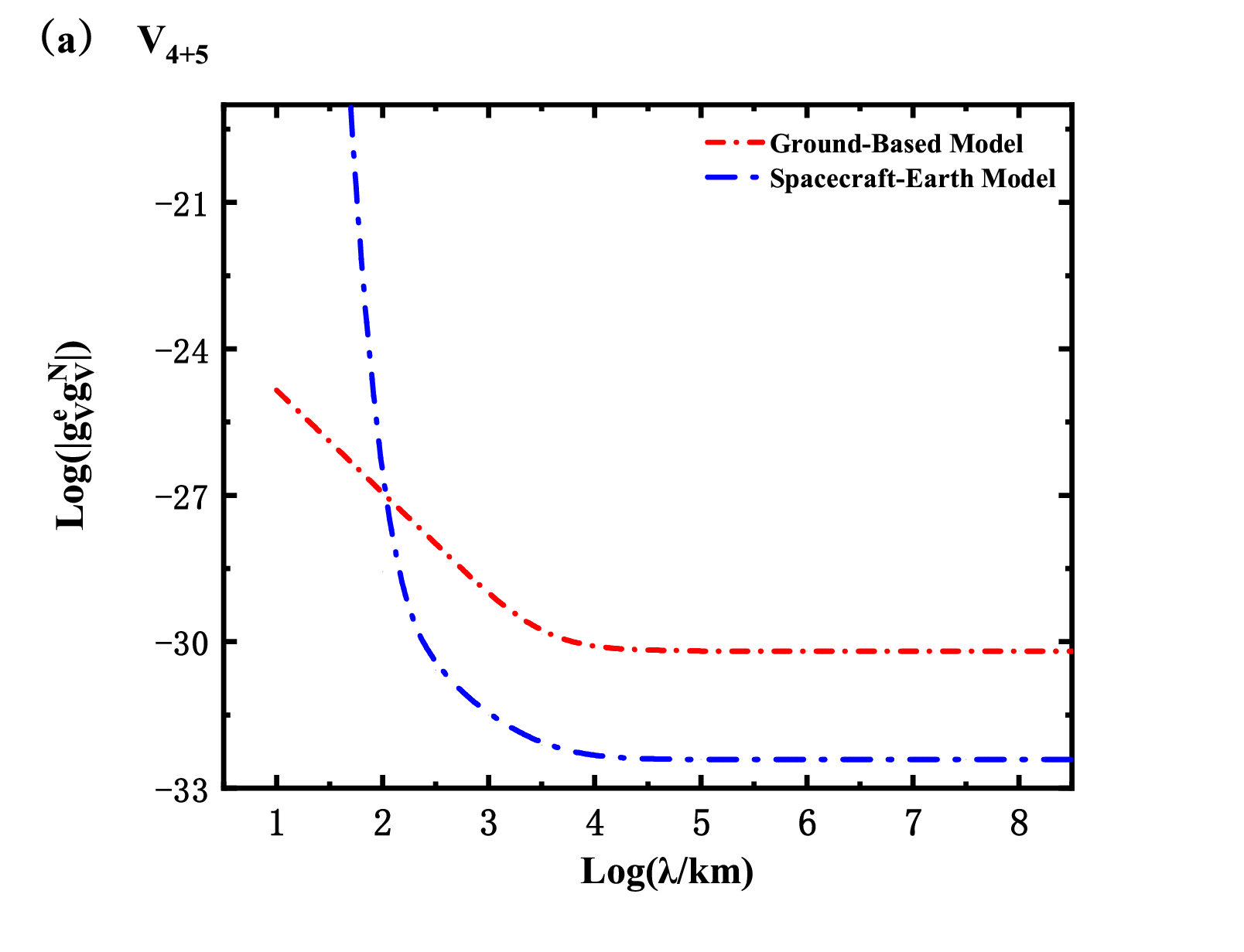}
  \end{subfigure}\hspace{-10mm}
  \begin{subfigure}{0.365\textwidth}
    \centering
    \includegraphics[width=\linewidth]{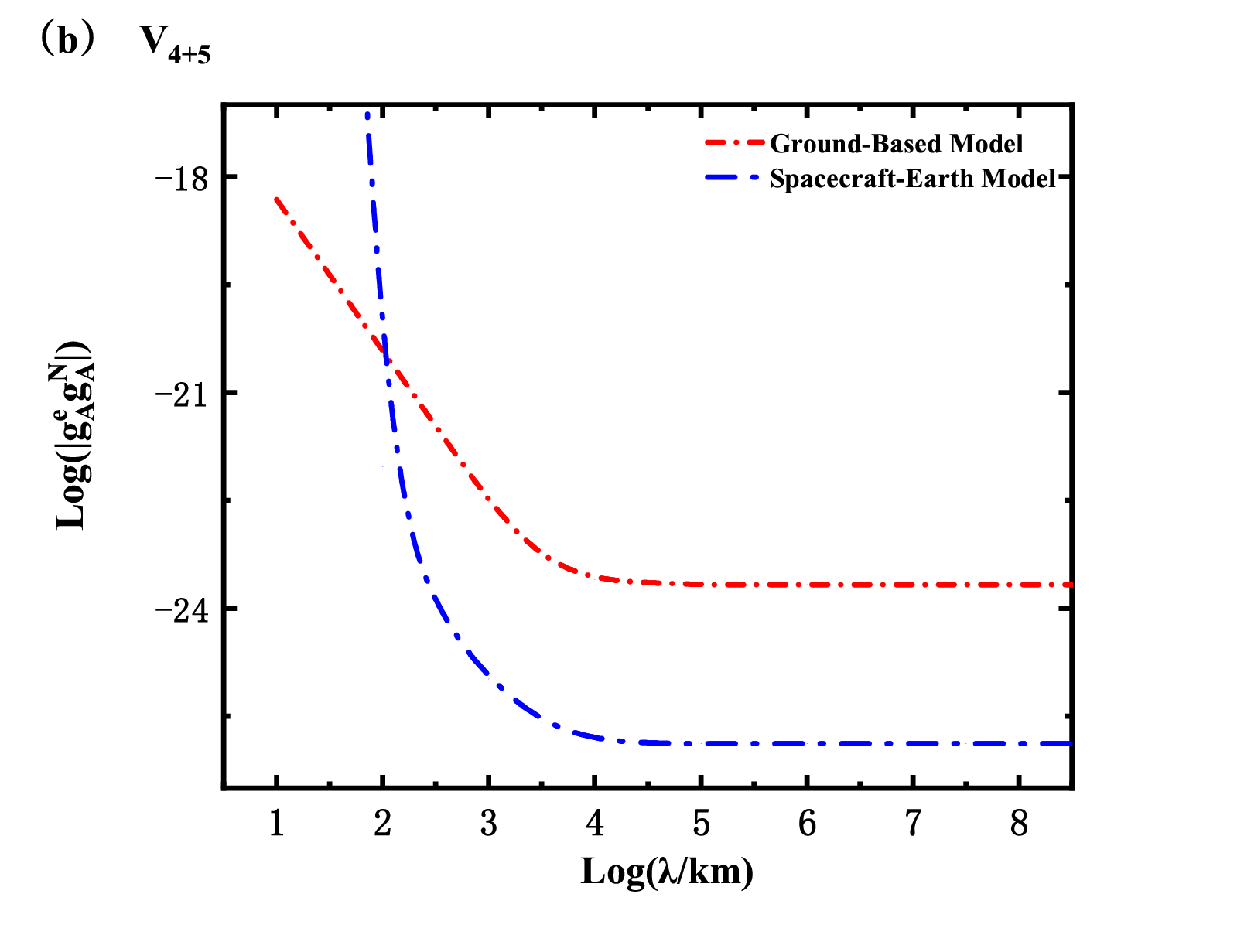}
  \end{subfigure}\hspace{-10mm}
  \begin{subfigure}{0.365\textwidth}
    \centering
    \includegraphics[width=\linewidth]{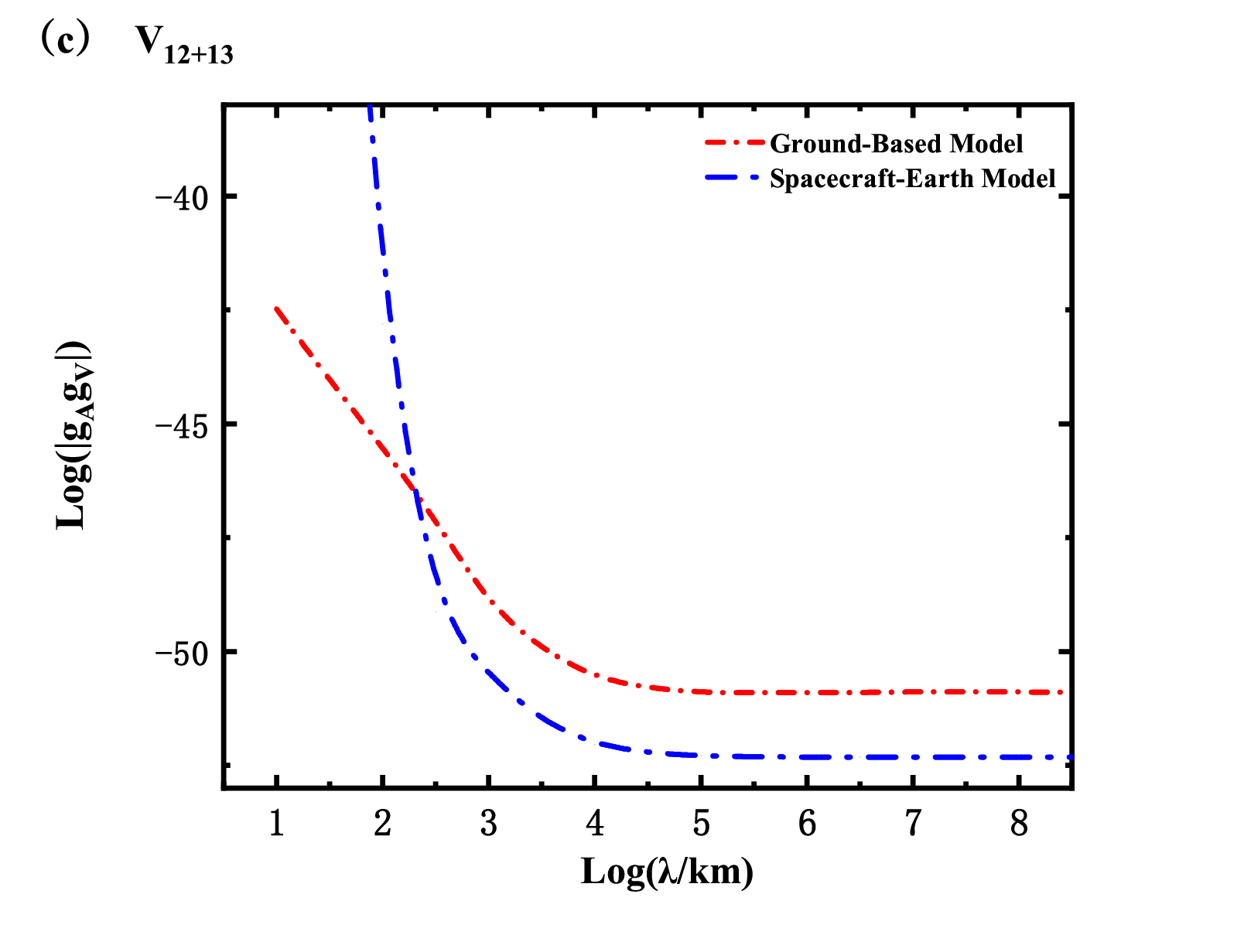}
  \end{subfigure}\hspace{-10mm}
  \caption{
  Expected bounds on long-range spin-velocity-dependent couplings for electron-nucleon ($e$-$N$) interactions. The red and green curves represent the expected constraints for an arbitrary energy-sensitivity condition, e.g., $10^{-20}$ eV. (a) (V-V)  couplings [Eq.~\ref{4+5VV}] for  $\left.V_{4+5}\right|_{V V}$. (b) (A-A)  couplings [Eq.~\ref{4+5AA}] for  $\left.V_{4+5}\right|_{A A}$. (c) (A-V) couplings [Eq.~\ref{12+13}] for  $V_{12+13}$.}
  
  \label{morepotentials}
\end{figure*}

The bounds on the Vector-Axial (V-A) couplings of $V_{6+7}$ (Fig.~\ref{V67RESULT}) are approximately 1.7 orders of magnitude more restrictive than those in the ground-based scheme for $\lambda = 10^{2.5}$ km, and at least 1.4 orders of magnitude more constrained in the flatter regions.

For $\lambda = 10^{2.5}$ km, the constraints on $V_{8}$, $V_{14}$ and $V_{16}$ (Fig.~\ref{bound}) are improved by up to 5, 3 and 4 orders of magnitude, respectively, while in the flatter parameter regions, the constraints are tightened by approximately three orders of magnitude across all three interactions. However, one should note that the $V_{8}(V_{16})$ constraints on the same coupling constants $g_Ag_A(g_Vg_A)$ are much weaker than those from $V_{14}(V_{6,7})$. The couplings for $V_{15}$ (Fig.~\ref{bound}(e)) are approximately three orders of magnitude more restrictive for $\lambda = 10^{3.5}$ km compared to the ground-based scheme, while as $\lambda$ increases, the improvement becomes less pronounced, indicating a more sensitive $\lambda$ dependence. In Ref.~\cite{38}, $V_{15}$ is found to be less reliable than other potentials, especially at short ranges,  because it is more susceptible to local inhomogeneities. Our ``Spacecraft-Earth" model can overcome this limitation and provide a more accurate constraint.

For the long-range spin-velocity dependent interactions for electron-nucleon ($e$-$N$) interactions, the improvement mainly appears where $\lambda$ exceeds the spacecraft altitude, as illustrated in Fig.~\ref{morepotentials}. Specially, the improvement is around 1.5 orders of magnitude in the flatter areas and up to 2.5 orders of magnitude for $\left.V_{4+5}\right|_{V V}$ and $\left.V_{4+5}\right|_{A A}$ with $\lambda = 10^{2.5}$ km. 
In comparison, it is 2 orders of magnitude for the odd-parity spin-velocity term ${V}_{12+13}$. 

{\it Summary.}---
In the present work, we proposed a novel theoretical conjecture to constrain further the coupling constants of long-range velocity-dependent interactions, commonly referred to as the fifth force. Drawing on Earth as an enormous source, we propose that the high-velocity orbital operations of a low-Earth-orbit spacecraft, whether satellites or space stations, can be exploited to overcome the limitations of ground-based experiments, such as fixed positions and slow speeds. Based on the spacecraft's periodic motion, we simulate the interaction strengths. Due to the much higher relative velocity, the bounds on most of the six velocity-dependent spin-spin interactions are expected to be at least three orders of magnitude more restrictive than earlier measurements, while the two spin-velocity terms are at least 1.5 orders of magnitude more restrictive. 

Previous works ~\cite{37,50} have highlighted the significance of the experimental location. The spacecraft's flexibility enables it to adapt over time, optimizing detection regions for the fifth force at varying velocities. The novel ``Spacecraft-Earth" model identifies optimal regions for the strongest signals, significantly improving current limits. Unlike previous analyses, its unique periodic line shapes allow distinguishing different interaction contributions. Given the diversity of dark matter hypotheses, this helps identify plausible candidates and types of interactions that may exist. Such periodicity also facilitates the extraction of signals from noise, thereby enhancing experimental accuracy.

Although the ``Spacecraft-Earth" model we proposed has the potential to significantly improve constraints on the coupling strength of spin-dependent interactions, conducting experiments in outer space remains a serious challenge. To achieve optimal sensitivity, it is necessary to address magnetic field shielding within the spacecraft, mitigate vibration-induced perturbations, and ensure the thermal stability of the environment in which the sensors operate. As a result, the sensor's actual detection sensitivity is likely not as good as that on the ground. Additionally, the carrier's inertial motion should be accounted for, and the specific experimental scheme should be carefully designed in conjunction with the spacecraft's attitude. On the other hand, impressive advancements in sensor sensitivity have been made, as reviewed in Ref.~\cite{7}. Thus, we are optimistic that the newly proposed ``Spacecraft-Earth" model will be of great help in making significant progress in detecting ultralight dark matter and tighten constraints on the exotic potentials mentioned above. 

{\it Acknowledgement.}---
This work is partly supported by the National Key R\&D Program of China under Grant No. 2023YFA16067003 and the National Science Foundation of China under Grant No. 12435007.

\end{document}